\documentclass[useAMS,usenatbib]{mnras}

\usepackage{mathrsfs}
\usepackage[T1]{fontenc}
\usepackage{ae,aecompl}

\newcommand\rxj{RXJ\,1131$-$1231}

\def\hst{\textit{HST}}

\def\kmsMpc {\rm km\,s^{-1}\,Mpc^{-1}}

\def\data{\boldsymbol{d}}
\def\datasca{d}
\def\PSF{\boldsymbol{w}}
\def\PSFsca{w}
\def\mapping{\bf{M}}
\def\lensing{\bf{L}}
\def\noise{\boldsymbol{n}}

\def\sersic{\boldsymbol{g}}
\def\H{H}
\def\s{\boldsymbol{s}}

\def\burring{\bf{K}}

\def\eda{\boldsymbol{\eta}}
\def\zda{\boldsymbol{\zeta}}
\def\tvec{\boldsymbol{t}}
\def\tsca{t}
\def\xivec{\boldsymbol{\nu}}

\def\tauvec{\boldsymbol{\xi}}

\def\ol{n}
\def\il{m}
\def\Tvec{{\bf T}}
\def\Tsca{\rm T}

\usepackage{natbib}
\usepackage{threeparttable}

\def\reff@jnl#1{{\rm#1\/}}
\def\aj{\reff@jnl{AJ}}                  
\def\araa{\reff@jnl{ARA\&A}}            
\def\apj{\reff@jnl{ApJ}}                
\def\apjl{\reff@jnl{ApJ}}               
\def\apjs{\reff@jnl{ApJS}}              
\def\apss{\reff@jnl{Ap\&SS}}            
\def\aap{\reff@jnl{A\&A}}               
\def\aapr{\reff@jnl{A\&A~Rev.}}         
\def\aaps{\reff@jnl{A\&AS}}             
\def\mnras{\reff@jnl{MNRAS}}            
\def\prd{\reff@jnl{Phys.Rev.D}}         
\def\prl{\reff@jnl{Phys.Rev.Lett}}      
\def\pasp{\reff@jnl{PASP}}              
\def\pasj{\reff@jnl{PASJ}}              
\def\nat{\reff@jnl{Nature}}             

\newcommand{\bd}{\begin{displaymath}}
\newcommand{\ed}{\end{displaymath}}
\newcommand{\be}{\begin{equation}}
\newcommand{\ee}{\end{equation}}
\newcommand{\beaa}{\begin{eqnarray*}}
\newcommand{\eeaa}{\end{eqnarray*}}
\newcommand{\bea}{\begin{eqnarray}}
\newcommand{\eea}{\end{eqnarray}}

\usepackage[dvips]{graphicx}	
\usepackage{amsmath}	
\usepackage{amssymb}	
\usepackage{gensymb}

\newcommand{\sref}[1]{Section~\ref{#1}}

\title[Strong lensing AO imaging with time delays for cosmography]{SHARP - \uppercase\expandafter{\romannumeral 3}: FIRST USE OF ADAPTIVE OPTICS IMAGING TO CONSTRAIN COSMOLOGY WITH GRAVITATIONAL LENS TIME DELAYS}

\author[G.C.-F.~Chen et al.]{
Geoff~C.-F.~Chen,$^{1,2,3}$ \thanks{E-mail: chfchen@ucdavis.edu}
Sherry~H.~Suyu,$^{1}$
Kenneth~C.~Wong,$^{1,4}$
\newauthor{
Christopher~D.~Fassnacht,$^{3}$
Tzihong~Chiueh,$^{2,5,6}$
Aleksi Halkola,
I~Shing~Hu,$^{7}$}
\newauthor{
Matthew W.~Auger,$^{8}$
Leon V.~E.~Koopmans,$^{9}$
David J.~Lagattuta,$^{10}$}
\newauthor{
John P.~McKean$^{9,11}$
and Simona Vegetti$^{12}$
}
\\
$^{1}$Institute of Astronomy and Astrophysics, Academia Sinica, P.O.~Box 23-141, Taipei 10617, Taiwan\\
$^{2}$Department of Physics, National Taiwan University, Taipei 10617, Taiwan\\
$^{3}$Department of Physics, University of California, Davis, CA 95616, USA\\
$^{4}$National Astronomical Observatory of Japan, 2-21-1 Osawa, Mitaka, Tokyo 181-8588, Japan \\
$^{5}$Institute of Astrophysics, National Taiwan University, Taipei 10617, Taiwan\\
$^{6}$Center for Theoretical Sciences, National Taiwan University, Taipei 10617, Taiwan\\
$^{7}$Department of Mathematics, National Taiwan University, Taipei 10617, Taiwan\\
$^{8}$Institute of Astronomy, University of Cambridge, Madingley Rd, Cambridge, CB3 0HA, UK\\
$^{9}$Kapteyn Astronomical Institute, University of Groningen, P.O.Box 800, 9700 AV Groningen, The Netherlands\\
$^{10}$CRAL, Observatoire de Lyon, Universit Lyon 1, 9 Avenue Ch. Andr, F-69561 Saint Genis Laval Cedex, France\\
$^{11}$Netherlands Institute for Radio Astronomy (ASTRON), P.O. Box 2, 7990 AA Dwingeloo, The Netherlands \\
$^{12}$Max Planck Institute for Astrophysics, Karl-Schwarzschild-Strasse 1, D-85740 Garching, Germany\\
}

\date{Accepted ---. Received ---; in original form \today}

\pubyear{2015}

\begin{document}
\label{firstpage}
\pagerange{\pageref{firstpage}--\pageref{lastpage}}
\maketitle

\begin{abstract}
Accurate and precise measurements of the Hubble constant are critical for 
testing our current standard cosmological model and revealing possibly new 
physics.  With \textit{Hubble Space Telescope} (\hst) imaging, 
each strong gravitational lens system with measured time delays can 
allow one to determine the Hubble constant with an uncertainty of $\sim$$7\%$.
Since \hst\ will not last forever, we explore adaptive-optics (AO) imaging as 
an alternative that can provide higher angular resolution than \hst\ imaging 
but has a less stable point spread function (PSF) due to atmospheric 
distortion. To make AO imaging useful for time-delay-lens cosmography, 
we develop a method to extract the unknown PSF directly from the
imaging of strongly lensed quasars.  In a blind test with two mock data sets 
created with different PSFs, we are able to recover the important cosmological 
parameters (time-delay distance, external shear, lens mass profile slope, and
total Einstein radius). Our analysis of the Keck AO image of the strong lens 
system \rxj\ shows that the important parameters for cosmography 
agree with those based on \hst\ imaging and modeling within 1-$\sigma$ 
uncertainties. Most importantly, the constraint on the model time-delay 
distance by using AO imaging with $0.045''$ resolution is tighter by 
$\sim$$50\%$ than the constraint of time-delay 
distance by using \hst\ imaging with $0.09''$ when a power-law mass
distribution for the lens system is adopted.  Our PSF reconstruction
technique is generic and applicable to data sets that have multiple
nearby point sources, enabling scientific studies that require
high-precision models of the PSF.
\end{abstract}




\section{Introduction}
The discovery of the accelerated expansion of the Universe
\citep{PerlmutterEtal99,RiessEtal98} and observations of the Cosmic
Microwave Background \citep[CMB; e.g.,][]{HinshawEtal12,AdeEtal15}  
have established a
standard cosmological paradigm where our Universe is spatially flat
and is dominated by cold dark matter (CDM) and dark energy: the
so-called flat $\Lambda$CDM model, where $\Lambda$ represents a
constant dark energy density.  While the CMB provides strong
constraints on the parameters of this model, a relaxation of the
assumptions in this model, such as spatial flatness or constant dark
energy density, leads to a strong degeneracy between the cosmological
parameters, particularly those with the Hubble constant $H_0$.
Therefore, independent and accurate measurements of $H_0$ provide one of
the most useful complements to the observations of the CMB in constraining
the spatial curvature of the Universe, dark energy equation of state,
and the number of neutrino species
\citep[e.g.,][]{Hu05,RiessEtal09,RiessEtal11,FreedmanEtal12,Suyuetal12b}. The
recent inferred value of Hubble constant $H_0=67.8\pm0.9\,\kmsMpc$,
based on the Planck satellite data of the CMB and the assumption of the
flat $\Lambda$CDM model, is low in comparison to several direct
measurements including those from the Cepheids distance ladder with
$H_0 = 74.3 \pm 1.5 {\rm (stat.)} \pm 2.1 {\rm (sys.)}\, \kmsMpc$
\citep{FreedmanEtal12} and $H_0 = 73.8 \pm 2.4\, \kmsMpc$
\citep{RiessEtal11}.  If this indication of tension is not
ruled out by systematic effects, then this could indicate new physics
beyond the standard flat $\Lambda$CDM model. Therefore, pinning down
the Hubble constant with independent methods is a key approach to
better understand our Universe.

Strong gravitational lensing with time delays provides a one-step
measurement of a cosmological distance in the Universe. The background
source is composed of a centrally varying source, such as an active
galactic nucleus (AGN), and its host galaxy. The time delays
between the multiple images of the source, induced
by the foreground lens, are given by 
$\Delta t$ $=\frac{1}{c}D_{\Delta t}\Delta\tau$.
Here, $\Delta\tau$ is dependent on the geometry and the gravitational 
potential of the lens system; $\Delta\tau$ can be tightly constrained by the
spatially extended images (we usually call them ``arcs'') of the lensed 
background galaxy \citep[e.g.,][]{KochanekEtal01, SuyuEtal09}, together with 
stellar kinematics of the foreground lens galaxy 
\citep[e.g.,][]{TreuKoopmans02, KoopmansEtal03, SuyuEtal10, SuyuEtal14}
and studies of the lens environment combined with ray-tracing through 
numerical simulations \citep[e.g.,][]{HilbertEtal07, HilbertEtal09, SuyuEtal10, FassnachtEtal11, GreeneEtal13, CollettEtal13}.  The stellar kinematics and 
lens environment studies are important for overcoming the mass-sheet 
degeneracy and source-position transformations in lensing 
\citep{FalcoEtal85, SchneiderSluse13, SchneiderSluse14, XuEtal15}.
Therefore, by measuring the time delays
between the multiple images and modeling the lens and line-of-sight mass 
distributions, we can constrain $D_{\Delta t}$, which is the so-called time-delay
distance that encompasses cosmological
dependences and is particularly sensitive to the Hubble constant
\citep[e.g.,][]{SuyuEtal10}.  The time delays in combination with the
stellar velocity dispersion measurements of the lens galaxy further
allow us to infer the angular diameter distance to the lens galaxy
\citep{ParaficzHjorth09, JeeEtal14}.

\citet{suyuEtal13} have shown that for each lens system we
can measure 
$H_0$ to $\sim$7$\%$ precision.
\textit{Hubble Space Telescope} (\hst) imaging is imperative for this
analysis because it not only provides high angular resolution but also
a stable point spread function (PSF) for the lens mass
modeling. However, \hst's lifetime is finite\footnote{And no equivalent 
optical space-based telescope might be forthcoming soon.}, 
and the angular resolution is also limited by its aperture size. 
Given the dozens of time-delay lenses from 
COSMOGRAIL\footnote{COSmological MOnitoring of
GRAvItational Lenses} \citep[e.g.][]{VuissozEtal07, VuissozEtal08, CourbinEtal11, TewesEtal13a, TewesEtal13b, RathnaEtal13, EulaersEtal13}, 
and hundreds of new lenses to be discovered 
in the near future \citep[e.g.,][]{OguriMarshall10, AgnelloEtal15,
  ChanEtal15, MarshallEtal15, MoreEtal15}, finding an alternative
long-term solution for this promising method is timely. 

One alternative approach is imaging from the ground via adaptive optics (AO), 
which is a technology used to improve the performance of optical systems by
reducing the effect of wavefront distortions \citep[e.g.,][]{RoussetEtal90,Beckers93,Watson97,Brase98}.
In other words, it aims
at correcting the deformations of an incoming wavefront by deforming a
mirror and thus compensating for the distortion. 
The advantages of using AO imaging are (1) the angular resolution obtained with
telescopes that are larger than \hst\ can be higher than that of \hst\
since a perfect AO system would lead to a diffraction limited PSF, (2)
ground-based telescopes are more accessible.
The disadvantage is that
we do not have a stable PSF model a priori, since the atmospheric distortion
varies both temporally and spatially across the image. Lens targets
typically do not have a nearby bright star within $\sim10$ arcseconds,
and stars at further angular distance from the target may be
insufficient in providing an accurate PSF model given the spatial
variation of the PSF across the field.

In \hst\ imaging, we can use the lensing arcs to constrain the lens
mass model by using the stable PSF of \hst\ to separate the arc from the
bright AGN, but we cannot do so in AO imaging. The contamination of
the AGN light on the lensing arcs in AO imaging makes it difficult to
constrain the lens model, and consequently $H_{0}$.  
One therefore needs to obtain a good PSF model for the AO data,
and there are recent studies that aim to do so directly from the AO imaging.
{\citet{Lagattuta10} use three Gaussian components as the 
PSF model to subtract the AGN light which is sufficient to study 
the lensing galaxy and its substructures. However, the analytical model is 
not sufficient to describe the complexity of the PSF 
\citep[see Figure 1 of][]{Lagattuta10} which could potentially impact the 
cosmographic measurements.
\citet{RusuEtal15} use either an analytic or a hybrid PSF to study the
host galaxies of the lensed AGNs \citep[see also][]{RusuEtal14}. The
hybrid PSF is built from elliptical Moffat profiles \citep[][]{Moffat69} with central parts
iteratively tuned to match a single AGN image.  While this hybrid PSF
is useful for extracting properties of the AGN host galaxy, the
central parts of the PSF model could manifest the noise pattern in the
image \citep[see Figure B.7 of][]{RusuEtal15} which also could potentially
impact cosmographic measurements. 
\citet{AgnelloEtal15} use an iterative method to reconstruct the PSF 
directly from lens imaging by averaging the doubly lensed AGN. 
This method is valid only when the lensed AGN are 
separated far enough from each other. 
For typical quad (four-image) lens systems, the
lensed AGNs are often close in separation (within $2''$), leading to
overlaps in the wings of the AGN images that are smeared by the PSF.
Our goal is to provide a general method
to overcome the unknown PSF model problem by extracting the PSF
directly from strong lensing imaging and simultaneously modeling the lens mass distribution.  
We test our method on simulated AO images,
and apply the method to the known gravitational lens \rxj\ with Keck
AO imaging, a part of data from SHARP\footnote{Strong-lensing High Angular Resolution Program 
(Fassnacht et al. in prep.)}, which is a 
project that focuses on studying known quadruple-image and Einstein ring 
lenses using high-resolution AO imaging, in order to probe their mass 
distributions in unprecedented detail \citep[e.g.,][Hsueh et al. submitted]{Lagattuta10, Lagattuta12, Vegetti12}.  
The gravitational lens system \rxj\ was discovered by \citet{SluseEtal03} who also measured the lens and source redshifts to be 0.295 and 0.658, respectively.
The \hst\ observations of the system \rxj\ have been modeled by \citet{suyuEtal13, SuyuEtal14} for cosmography and more recently by \citet{BirrerEtal15}.

The outline of the paper is as follows. In Section \ref{sec:obs}, we
describe the observation of \rxj\ with the adaptive optics imaging
system at the Keck Observatory.  We briefly recap in \sref{sec:theory}
the basics of cosmography with time-delay lenses.  In Section
\ref{sec:method}, we describe our new procedure to analyze AO images
without information on the PSF in advance. In Section
\ref{sec:blindtest}, we use simulated data to test and verify the
method. In Section \ref{sec:realdata}, we demonstrate the results
from real data and provide a
comparison between the results from \hst\ imaging and AO imaging.  
Finally, we summarize in Section \ref{sec:summary}.

\begin{figure}
\includegraphics*[scale=0.58]{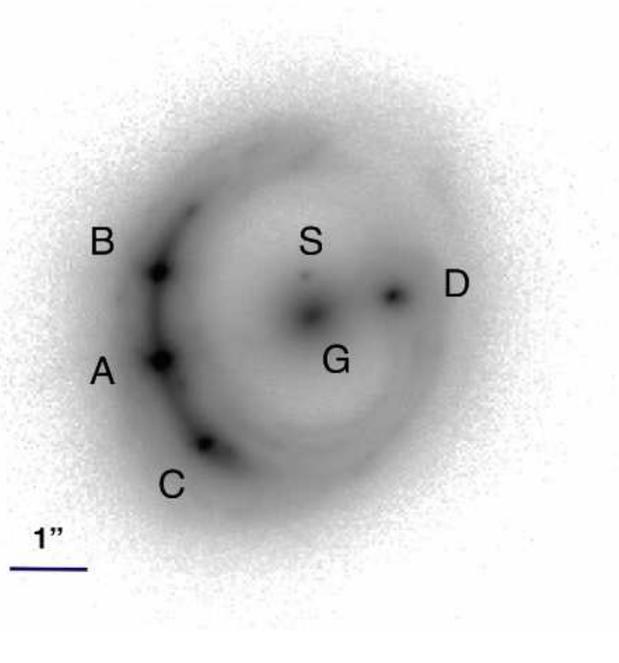}
\caption{Keck AO image (K$^\prime$ band) of the gravitational lens RXJ1131-1231. The lensed AGN image  of the spiral source galaxy are marked by A, B, C and D, and the star-forming regions in the background spiral galaxy form plentiful lensed features. The foreground main lens and the satellite are indicated by G and S, respectively.}
\label{fig:image}
\end{figure}

\section{OBSERVATION} 
\label{sec:obs}

The \rxj\ system was observed on the nights of UT 2012 May 16 and
May 18 with the Near Infrared Camera 2 (NIRC2) on the Keck-2
Telescope \citep[e.g.,][]{wizinowich03}. This image was a part of SHARP data.
The adaptive optics corrections were achieved through the
use of a $R = 15.8$  tip-tilt star located 54.5 arcseconds from the lens
system and a laser guide star.  The system was observed in the ``Wide
Camera'' mode, which provides a roughly $40^{\prime\prime} \times
40^{\prime\prime}$ field of view and a pixel scale of 0.0397 arcseconds.  
This pixel scale slightly undersamples the point spread function (PSF), but
the angular extent of the lens system and the distance from the
tip-tilt star made the use of the Wide Camera the preferable approach.

The observations consisted of 61 exposures, each consisting of 6
coadded 10~s exposures, for a total on-source integration time of
3660~s.  The data were reduced by a python-based pipeline that has
steps that do the flat-field correction, subtract the sky, correct for
the optical distortions in the raw images, and combine the calibrated
data frames \citep[for details, see][]{auger_eels}.  The final image has
a pixel-scale of 0.04~arcseconds and is shown in Figure \ref{fig:image}.

\section{BASIC THEORY}
\label{sec:theory}

\subsection{The Theory of Gravitational Lensing \\with Time Delay}
\label{sec:theory:timedelays}

In this section we briefly explain the relation between gravitational 
time delays and cosmology. 
When a light ray passes near a massive object, it
experiences a deflection in its trajectory and acquires a time delay
by the gravitational field with respect to the travel time without the
massive object.
Therefore, the time delay has two contributions: (1) the geometric delay, 
$\Delta t_{\rm geom}$, which is caused by the bent trajectory
being longer than the
unbent one, and (2) the gravitational delay, $\Delta t_{\rm grav}$, 
which is due to the fact that the space and time are affected around the 
gravitational field, so after integrating the gravitational potential along 
the path, a far away observer receives the light later by a Shapiro delay 
\citep{Shapiro64, Refsdal64}. 

The combination of the two delays is

\begin{equation}
\label{eq:theory6}
\Delta t=\frac{D_{\Delta t}}{c}\left[\frac{1}{2}(\boldsymbol{\theta}-\boldsymbol{\beta})^{2}-\psi(\boldsymbol{\theta})\right],
\end{equation}
where $\boldsymbol{\theta}$, $\boldsymbol{\beta}$, and
$\psi(\boldsymbol{\theta})$ are the image coordinates, the source
coordinates, and the lens potential respectively. The time-delay distance
is defined as
\begin{equation}
\label{eq:theory7}
D_{\Delta t}\equiv(1+\textit{z}_{\rm d})\frac{D_{\rm d}D_{\rm s}}{{D_{\rm ds}}}\propto H_{0}^{-1},
\end{equation}
where $D_{\rm d}$,
$D_{\rm s}$ and $D_{\rm ds}$ are the angular diameter distances to
the lens, to the source, and between the lens and the source,
respectively. 
Thus, we can measure $D_{\Delta t}$ via gravitational lensing with time delays. 
Notice that the gradient of the term in the square brackets in 
Equation (\ref{eq:theory6}) vanishes at the positions of the lensed images 
and yields the lens equation
\be
\boldsymbol{\beta} = \boldsymbol{\theta} -
\nabla\psi(\boldsymbol{\theta}),
\ee
which governs the deflection of light rays.

We refer the reader to, e.g.,
\citet{SchneiderEtal06}, \citet{Bartelmann10}, \citet{Treu10},
\citet{SuyuEtal10}, \citet{Treu14} for more details. 

\begin{figure*}
\includegraphics*[scale=0.7]{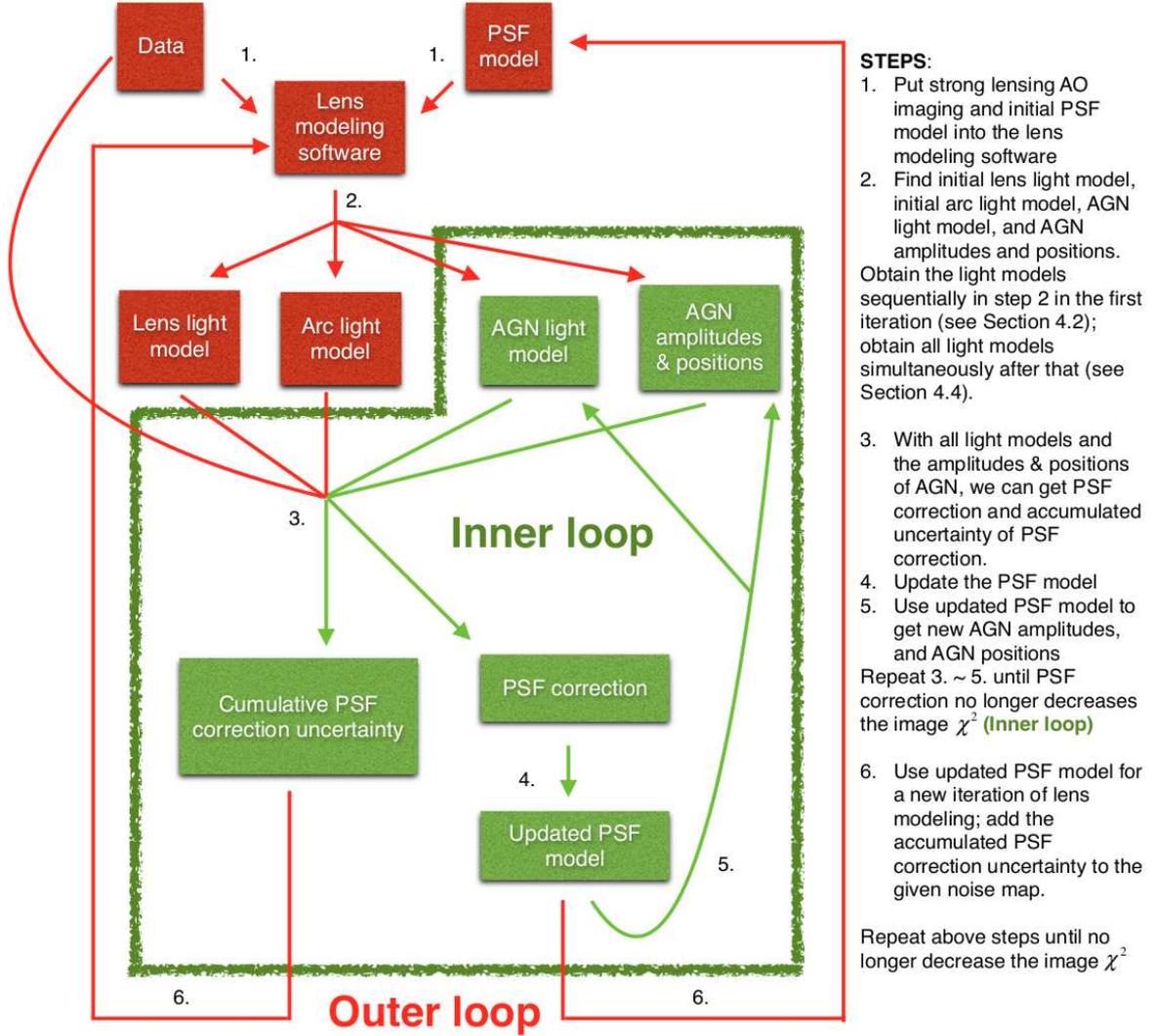}
\caption{The flow-chart describes the overall procedures in Section
  {$\ref{sec:method}$}. We use the procedures to reconstruct the
  PSF directly from lens image and do the lens modeling. In step 1, we
  use a nearby star (or one of the lensed AGN itself) as the initial
  PSF; in step 2, we sequentially obtain the lens light, arc light, AGNs 
  light, and the positions and relative amplitudes of AGNs; steps 3 to 5
  form an inner loop to add the correction (fine structures) into the PSF and 
  accumulate the correction uncertainties; 
  in step 6, we enter the outer loop which updates the image covariance matrix,
  PSF of all light model, and then repeat the full procedure until the
  image $\chi^2$ no longer decreases.
}
\label{fig:diagram}
\end{figure*}

\subsection{Probability Theory}
\label{sec:theory:probability}

A meaningful measurement should have an uncertainty as a reference
and it is also the key to confirm or rule out possible models.
Thus, we need to
analyze our data based on a probability theory that can present this
idea. Bayes' theorem provides the conditional probability
distribution, so we can obtain the posterior probability distribution 
of the model parameters given the data from Bayes' rule. For
example, if we are interested in the posterior of the parameters
$\boldsymbol{\cal \pi}$ of the hypothesis model $\H$ given the data
$\data$, it can be expressed as

\begin{equation}
\label{basic_bayes_rule}
\overbrace{P(\boldsymbol{\cal \pi}|\data,\H)}^{\text{posterior}}=\frac{\overbrace{P(\data|\boldsymbol{\cal \pi},\H)}^{\text{likelihood}}\overbrace{P(\boldsymbol{\cal \pi}|\H)}^{\text{prior}}}{\underbrace{P(\data|\H)}_{\text{evidence
      (marginalized likelihood)}}},
\end{equation}
where the Bayesian evidence can be used to rank the model and our
prior based on the data  
\citep[e.g.,][]{Mackay92, HobsonEtal02,MarshallEtal02}

In addition, if we are interested in the posterior of a specific
parameter, ${\cal \pi}_{N}$, the posterior distribution can be 
obtained by marginalizing over other parameters

\begin{equation}
\label{basic_bayes_rule1}
P({\cal \pi}_{N}|\data,\H)=\int P(\boldsymbol{\cal \pi}|\data,\H) \prod_{i=1}^{N-1}{d{\cal \pi}_{i}}.
\end{equation}

\begin{figure*}
\centering
\includegraphics[scale=0.65]{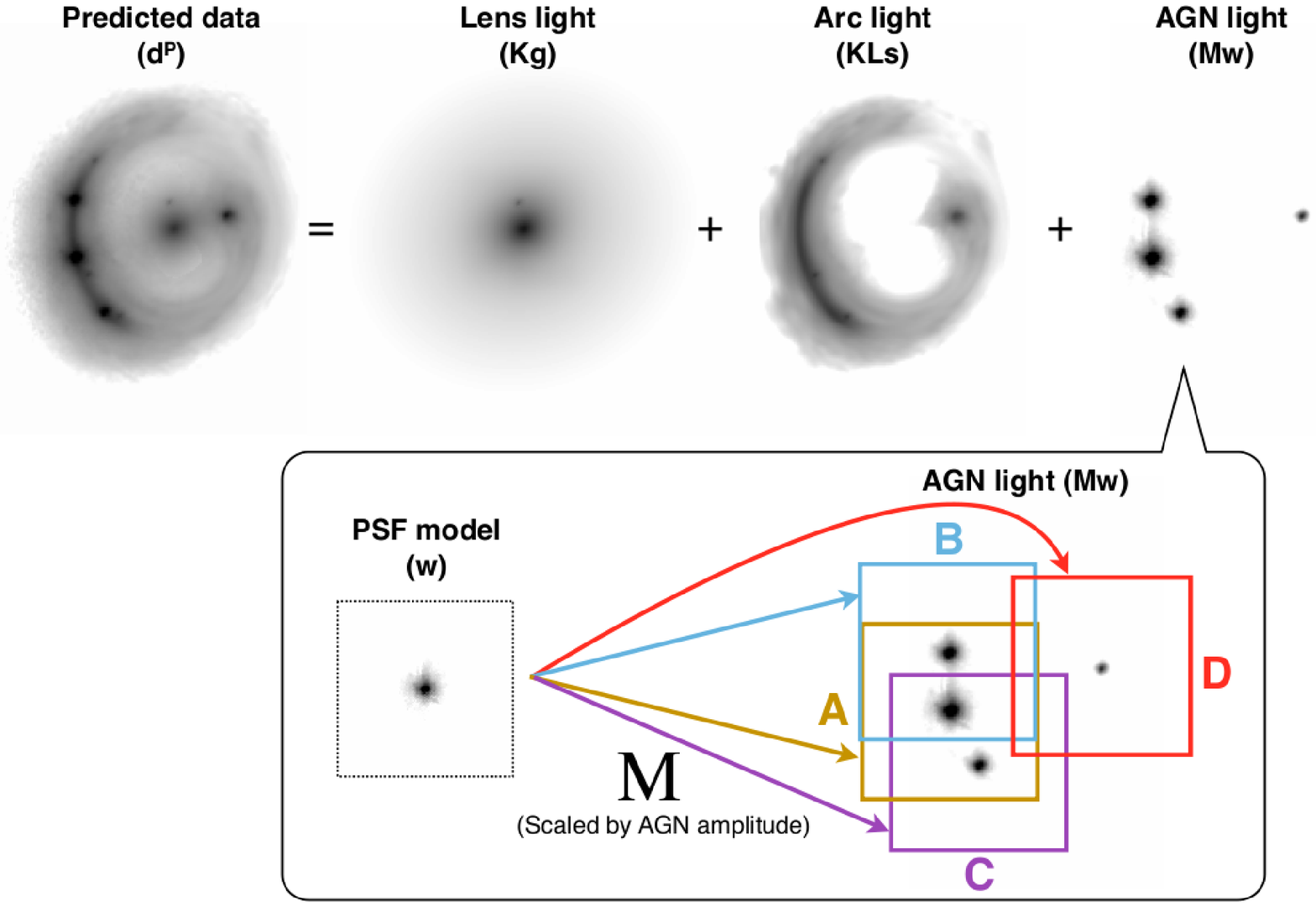}
\caption{
Top panel: we decompose the image
into lens light, arc light, and AGN light sequentially. 
Bottom panel: we model the AGN light by placing the PSF grid
(described by vector $w$) at each of the AGN positions and scaling
each PSF by its respective AGN amplitude.  This procedure can be
characterized by a matrix $\mapping$, such that the AGN light model on
the image plane can be expressed as $\mapping \PSF$.}
\label{fig:decom}
\end{figure*}

\subsection{Markov chain Monte Carlo}
\label{sec:theory:mcmc}

Obtaining the probability distribution function of the parameters in a
model can be non-trivial, especially when the number of parameters
is high.  It is computationally unfeasible to explore a high-dimensional 
parameter space on a regular grid since the number 
of the grid points for the task exponentially increases with the number of dimensions. 
Due to the fact that the parameter space is typically large in strong lensing 
analyses, one can bypass the use of grids by obtaining samples in the
multi-dimensional parameter space that represent the probability
distribution (i.e., the number density of the samples is proportional
to the probability density).  A Markov Chain Monte Carlo (MCMC)
provides an efficient way to draw samples from the posterior
probability density function (PDF) of the lens parameters, because of
the approximately linear relation between the computational time and
the dimension of the parameter space.  

We use MCMC sampling that is implemented in {\sc Glee}, a strong
lens modeling software developed by S.~H.~Suyu and A.~Halkola
\citep{SuyuHalkola10,SuyuEtal12a}. It is based on Bayes' theorem and 
follows \citet{DunkleyEtal05} to achieve efficient sampling and to test  
convergence. The pragmatic procedure for convergence is described in 
\citet{SuyuHalkola10}. We use Bayesian language in the following sections.

\section{Method: PSF reconstruction and lens modeling}
\label{sec:method}

In this section, we describe a novel procedure to analyze the AO imaging
without a PSF model a priori. 
Readers who are not planning to use this method may
wish to proceed directly to Section \ref{sec:blindtest} on the
scientific results enabled by the method.

The assumption of this strategy is that
the PSF does not change significantly within several arcseconds, which
is typically valid in AO imaging \citep{van06,wizinowich06}. 
We show an overall flow chart in
Figure \ref{fig:diagram} to illustrate how to obtain iteratively the
PSF, background source intensity, the lens mass
and light model. 

In Section \ref{sec:method:notation}, we decompose the observed light
from the lens system into three components (lens galaxy, lensed arcs
of the background source galaxy, and the lensed background AGN) and 
introduce the notation that we will use in the subsequent discussion.  
In Section \ref{sec:method:initPSF}, we obtain the preliminary global
structure of AGN light model, while separating the lens light and arc
light. In Section \ref{sec:method:iterative}, we obtain the fine
structure of the AGN light and incorporate it into the preliminary AGN light
model. This is accomplished by correcting the PSF model. In \sref{sec:method:lensModelUpdated}, 
we update the PSF and use it to model the lens mass and source intensity distributions.  
Since the lens galaxy light is quite smooth and less sensitive to the PSF model, 
we use the PSF built from the AGN light for the lens galaxy light model.
The PSF updating and lens mass modeling are repeated until the corrections to the PSF become 
insignificant. (See the criteria in Section \ref{sec:method:innerloopstop} and Section \ref{sec:method:outerloopstop}.)

\subsection{Light components of the lens system}
\label{sec:method:notation}

As shown in Figure \ref{fig:decom}, our model for the observed light
in the lens system on the image plane has three contributions: the
lens galaxy light, the arc light (the lensed background source, i.e.,
the host galaxy of the AGN), and the light of the multiple AGNs on the
image plane.  We define
\begin{equation}
\label{fundamental}
\data=\data^{\text{P}}+\noise,
\end{equation}
where $\data$ is the vector of observed data (image pixel values),
\begin{equation}
\label{generaleq}
\data^{\text{P}}=\overbrace{{\burring}\sersic}^{\text{lens light}}+\overbrace{{\burring}{\lensing}\s}^{\text{arc light}}+\overbrace{{\mapping}\PSF}^{\text{AGN light}},
\end{equation}
and $\noise$ is the noise in the data characterized by the covariance
matrix $\bf{C}_{\text{D}}$ (we use subscript D to indicate
``data''). The blurring matrix ${\burring}$ accounts for the PSF
convolution, the vector $\sersic$ is the image pixel 
values of the lens galaxy light, the matrix $\lensing$ maps source intensity 
to the image plane, the vector $\s$ describes the source surface brightness 
on a grid of pixels, the matrix $\mapping$ is composed using
the positions and the intensities of the AGNs, and $\PSF$ is the vector 
of pixel values of the PSF grid. We refer to \citet{TreuKoopmans04} for 
constructing $\burring$ and $\lensing$, and 
illustrate the effect of $\mapping$ in Figure \ref{fig:decom}.

At first, since we do not know the 
AO PSF a priori, 
$\burring$ and $\PSF$ are just the initial blurring matrix and PSF
grid values, respectively. 
As we iteratively model the light components and correct the PSF, 
we update $\PSF$ (and subsequently ${\burring}$).

\subsection{Determining the light components}
\label{sec:method:initPSF}

The goal in this section is to obtain the preliminary model of each of
the three light components. In step 1 of Figure \ref{fig:diagram}, we
input the observed image into the lens modeling software {\sc Glee} with 
a nearby star as our initial PSF model.  If there is no nearby star, 
any star in the field can be used as the initial PSF or we can use one of the AGN images.  A different initial PSF does not affect the final results, 
although we note that a good initial PSF would be helpful as they would require fewer iterations of PSF corrections.
In step 2, we decompose the
predicted total light sequentially into lens light, arc light, and AGN
light.  We detail this process  
in Section
\ref{sec:method:initPSF:lenslight} to Section
\ref{sec:method:initPSF:AGNlight} below.

\subsubsection{Lens Light Model (Step 2)}
\label{sec:method:initPSF:lenslight}
For modeling the light distribution of the lens galaxy, we use
parametrized profiles, such as the elliptical S$\acute{\text{e}}$rsic profile,
\begin{displaymath}
I_{\rm S}(\theta_{1},\theta_{2})\qquad \qquad \qquad \qquad \qquad \qquad \qquad \qquad \qquad \qquad 
\end{displaymath}
\begin{equation}
\label{eq:lenslight0}
\quad=I_{\rm s}\,\mathrm{exp}\left[-k\left(\left(\frac{\sqrt{\theta_{1}^{2}+\theta_{2}^{2}/q_{\text{L}}^{2}}}{R_{\text{eff}}}\right)^{1/n_{\text{s}\acute{\text{e}}\text{rsic}}}-1\right)\right],
\end{equation}
where $I_{\rm s}$ is the amplitude, $k$ is a constant such that
$R_{\text{eff}}$ is the effective radius, $q_{\text{L}}$ is the minor-to-major axis ratio, and $n_{\text{s}\acute{\text{e}}\text{rsic}}$ is the
S$\acute{\text{e}}$rsic index \citep{Sersic68}.  

In order to get a preliminary model of the lens light, we mask out the
arc light and AGN light region; that is, we boost the uncertainty of
the region where the arc light and the AGN light are apparently
dominant. Thus, in the fitting region, equation (\ref{generaleq})
becomes effectively  
\begin{equation}
\label{eq:lenslight0.5}
\data^{\text{P}}=\burring\sersic.
\end{equation}
By Bayes' rule, we have
\begin{equation}
\label{eq:lenslight1}
{P(\eda|\data)}\propto {P(\data|\eda)}{P(\eda)},
\end{equation}
where $\eda$ represents the parameters of lens light (such as $I_{s},
q_{\text{L}},n_{\text{s}\acute{\text{e}}\text{rsic}},R_{\text{eff}}$). 
We assume uniform prior on the lens light parameters, so we want to obtain
\begin{equation}
\label{eq:lenslight2}
{P(\data|\eda)} = \frac{\mathrm{exp}[-E_{\text{D,mArcAGN}}(\data|\eda)]}{Z_{\text{D,mArcAGN}}},
\end{equation}
where,
\begin{displaymath}
{E_{\text{D,mArcAGN}}}(\data|\eda)\qquad\qquad\qquad\qquad\qquad\qquad
\end{displaymath}
\begin{displaymath}
=\frac{1}{2}({\data-\burring\sersic})^{\text{T}}{\bf C}_{\text{D,mArcAGN}}^{-1}({\data-\burring\sersic})
\end{displaymath}
\begin{equation}
\label{eq:lenslight3}
=\frac{1}{2}\chi^{2}_{\rm mArcAGN}, \qquad\qquad\qquad\qquad\quad\,\,\,
\end{equation}
and 
\begin{equation}
\label{eq:lenslight4}
Z_{\text{D,mArcAGN}}=(2\pi)^{N_{\rm \datasca}/2}(\text{det }{\bf C}_{\text{D,mArcAGN}})^{1/2}
\end{equation}
is the  normalization for the probability. The covariance matrix,
${\bf C}_{\text{D,mArcAGN}}$, is the original covariance matrix
with entries corresponding to the arc and AGN mask region boosted (see
Appendix \ref{app:a}), and $N_{\rm \datasca}$ is the number of
image pixels. We denote $\hat{\eda}$ as the maximum likelihood
parameters (which maximizes equation \ref{eq:lenslight2}).

Since the initial PSF is a prototype, usually
there are some significant residuals in the lens light center when maximizing
the posterior of lens light parameters. However, this does not affect the
subsequent lens light prediction in the arc region, because the residuals
are far from the arc regions.  To recap, we can obtain $\hat{\eda}$ by 
masking out the arc light and AGN light regions.

\begin{figure*}
\centering
\includegraphics[scale=0.5]{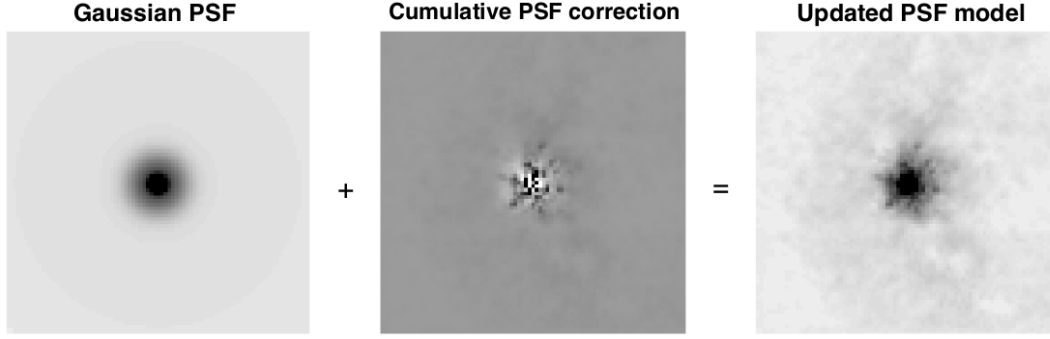}
\caption{The left panel is the global structure of the PSF. 
The middle panel is the cumulative fine structure of the PSF. 
We add the fine structure to global structure to get the PSF model 
in the right panel.}
\label{fig:corr}
\end{figure*}

\subsubsection{Arc Light Model (Step 2)}
\label{sec:method:initPSF:arclight}
For modeling the arc light, we describe the source intensity on a grid
of pixels on the source plane and map the source intensity values onto
the image plane using a lens mass model (via the operation
$\burring\lensing\s$ in equation (\ref{generaleq})).  We use 
elliptically symmetric power-law distributions to model the
dimensionless surface mass density of lens galaxies,
\begin{equation}
\label{eq:arclight-1}
\kappa_{\text{pl}}(\theta_{1},\theta_{2})=\frac{3-\gamma'}{1+q}\left( \frac{\theta_{\text{E}}}{\sqrt{\theta_{1}^{2}+\theta_{2}^{2}/q^2}}   \right)^{\gamma'-1},
\end{equation}
where $\gamma'$ is the radial power-law slope ($\gamma'=2$
corresponding to isothermal), $\theta_{\rm E}$ is the Einstein radius, and
$q$ is the axis ratio of the elliptical isodensity contour. In
addition to the lens galaxies, we include a constant external shear with
the following lens potential in polar coordinates $\theta$
and $\varphi$: 
\begin{equation}
\label{eq:arclight0}
\psi_{\text{ext}}(\theta, \varphi)=\frac{1}{2}\gamma_{\text{ext}}\theta^{2}\cos2(\varphi-\phi_{\text{ext}}),
\end{equation}
where $\gamma_{\text{ext}}$ is shear strength and $\phi_{\text{ext}}$
is the shear angle. The shear position angle of
$\phi_{\text{ext}}=0^{\circ}$ corresponds to a shearing along
$\theta_{1}$ whereas $\phi_{\text{ext}}=90^{\circ}$ corresponds to
shearing along $\theta_{2}$.\footnote{Our (right-handed) coordinate
  system $(\theta_{1},\theta_2)$ has $\theta_1$ along the East-West
  direction and $\theta_2$ along the North-South direction.}

We model the arc light with the lens light fixed. Since the AGN light
dominates near the AGN image positions, we mask
out the region where the arc is hard to be seen; that is, we want to
minimize the contribution to the source intensity reconstruction from
the AGN light.  Since the regions of the AGN are masked out, we
temporarily\footnote{We will put the AGN component back in next section, \ref{sec:method:initPSF:AGNlight}} drop the AGN component, $\mapping\PSF$, in Equation
(\ref{generaleq}), which given $\hat{\eda}$ becomes 
\begin{equation}
\label{eq:arclight10}
\data^{\text{P}}={\burring}\hat{\sersic}+\burring\lensing\s,
\end{equation}
where $\hat{\sersic}=\sersic(\hat{\eda})$.
The posterior based on the arc light is 
\begin{equation}
\label{eq:arclight1}
P({\zda}|{\data},{\Delta\tvec},{\hat{\eda}})\propto {P(\data,\Delta\tvec|\hat{\eda},\zda)}{P(\zda)},
\end{equation}
where $\zda$ are the parameters of the lens mass distributions (such as $\gamma'$, $\theta_{\text{E}}$, $\gamma_{\text{ext}}$).
The likelihood of the data can be expressed as
\begin{equation}
\label{eq:arclight2}
P({\data},{\Delta\tvec}|{\hat{\eda},\zda})= \int\rm{d}{\s}\:{P({\data},{\Delta\tvec}|{\hat{\eda}},{\zda},{\s})}{P(\s)},
\end{equation}
where
\begin{displaymath}
{P(\data,\Delta\tvec|\hat{\eda},\zda,\s)} \qquad \qquad \qquad \qquad \qquad \qquad \qquad
\end{displaymath}

\begin{displaymath}
=\frac{\mathrm{exp}[-E_{\text{D,mAGN}}(\data|\hat{\eda},\zda,\s)]}{Z_{\text{D,mAGN}}}\qquad \qquad \quad
\end{displaymath}

\begin{displaymath}
\label{eq:arclight3}
\cdot\prod_{i=1}^{N_{\text{AGN}}}\frac{1}{\sqrt{2\pi}\sigma_{{\rm
      AGN},i}}\mathrm{exp}\left(-\frac{|\boldsymbol{\theta}_{{\rm
        AGN},i}-\boldsymbol{\theta}_{{\rm AGN},i}^{\rm
      p}|^{2}}{2\sigma_{{\rm AGN},i}^{2}}\right)
\end{displaymath}

\begin{equation}
\label{eq:arclight4}
\qquad \qquad
\cdot\prod_{i=1}\frac{1}{\sqrt{2\pi}\sigma_{\Delta\tsca,i}}\mathrm{exp}\left[-\frac{(\Delta\tsca_{i}-\Delta\tsca_{i}^{\rm
      p})^{2}}{2\sigma_{\Delta\tsca,i}^{2}}\right],
\end{equation}

\begin{displaymath}
{E_{\text{D,mAGN}}(\data|\hat{\eda},\zda,\s)} \qquad \qquad\qquad\qquad\qquad\qquad\qquad\qquad\qquad
\end{displaymath}
\begin{equation}
\label{eq:arclight5}
=\frac{1}{2}({\data-{\burring}\hat{\sersic}-\burring\lensing\s})^{\text{T}}{\bf C}_{\text{D,mAGN}}^{-1}({\data-{\burring}\hat{\sersic}-\burring\lensing\s}),
\end{equation}
and
\begin{equation}
Z_{\text{D,mAGN}}=(2\pi)^{N_{\rm \datasca}/2}(\text{det }{\bf C}_{\text{D,mAGN}})^{1/2}
\end{equation}
is the normalization for the probability. We discuss the ``mAGN'' regions 
in Appendix \ref{app:a}. In the second term of Equation (\ref{eq:arclight4}),
$\boldsymbol{\theta}_{{\rm AGN},i}$ is the measured AGN image position
and ${\sigma}_{{\rm AGN},i}$
is the estimated positional uncertainty of AGN image $i$; in the third term, $\Delta\tsca_{i}$ is the measured time delay with uncertainty 
$\sigma_{\Delta\tsca,i}$ for image pair $i=AB,CB,$ or $DB$.  
After we maximize the likelihood of the data, we
obtain $\hat{\zda}$, and also the predicted arc light of the
reconstructed background source intensity, $\hat{\s}$, of the AGN host galaxy.
Note that if there is no time-delay information, one can remove the last term 
in Equation (\ref{eq:arclight4}).

\begin{figure*}
\centering
\includegraphics[scale=0.4]{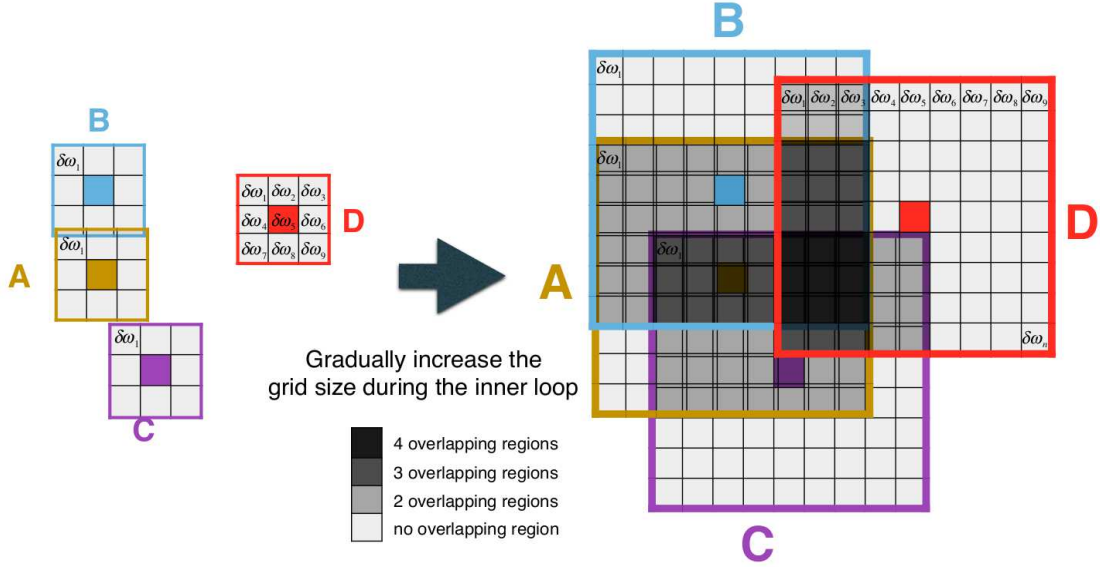}
\caption{
The PSF correction grids of the iterative PSF correction scheme.  In
the inner loop of the PSF correction scheme (same $\ol$ but different
$\il$), we start with a small correction grid $\delta\PSF$ and
increase it sequentially.  This accommodates for the larger
corrections needed in the central parts of the AGN.  Left panel: a
small PSF correction grid is placed at each of the four AGN images A,
B, C and D in Figure \ref{fig:image} via the matrix $\mapping$, and
the values of the PSF correction grid is determined via a linear
inversion to reduce the overall image residuals.  Since the AGN
centroids are typically non-integral pixel values, we linearly
interpolate the correction grid onto the image plane.  Right panel:
the enlarged corrections grids after several iterations of PSF
correction, showing overlap between the grids.  When the peripheral
area of a correction grid overlaps with the central parts of another
AGN image (e.g., AGN image C in the lower-right parts of the correction
grid of image A), we mask out the center of the AGN region in order to 
prevent the correction grids from absorbing the residuals which come from the 
mismatch of the sharp intensity of AGN center (see Appendix \ref{app:a} for more details). 
}
\label{fig:pixgrid_overlap}
\end{figure*}

\subsubsection{AGN Light Model (Step 2)}
\label{sec:method:initPSF:AGNlight}
In Equation (\ref{generaleq}), we use $\mapping\PSF$ to represent the 
AGN light. In the next section, we further decompose the PSF, $\PSF$, 
into the global structure and the fine structure that are shown in 
Figure \ref{fig:corr}.  In particular, we define
\begin{equation}
\label{eq:AGNlight1}
\PSF=\PSF_{[0]}+\Tvec_{[0]}\delta\PSF_{[0]},
\end{equation}
where $\PSF_{[0]}$ is the vector of global structure, $\delta\PSF_{[0]}$ 
is the vector of fine structure and the subscript, ${[0]}$, represents the zero-th iteration. Since, in this section, we focus on the global 
structure of the PSF, we postpone the discussion of $\Tvec$ to Equation (\ref{eq:fine_structure0}) and let
\begin{equation}
\label{eq:AGNlight6}
\PSF=\PSF_{[0]}.
\end{equation}
By using $\hat{\eda}$, $\hat{\zda}$, and $\hat{\s}$ from the previous 
two sections and keeping them fixed, we model the global structure of 
the PSF with Gaussian profiles, each of the form
\begin{equation}
\label{eq:AGNlight3}
I_{\rm G}(\theta_{1},\theta_{2})=I_{\text{g}}\,\mathrm{exp}\left[-\frac{\theta_{1}^{2}+(\theta_{2}^{2}/q_{\text{g}}^{2})}{2\sigma_{\text{g}}^{2}}\right],
\end{equation}
where $I_{\text{g}}$ is the amplitude, $q_{\text{g}}$ is the axis ratio, 
and $\sigma_{\text{g}}$ is the width.
We find that a few Gaussians ($\sim2-4$) with a common centroid are 
sufficient in describing the global structure of the PSF.\footnote{The different Gaussian components can vary their amplitudes, position angles and axis ratios.}
Substituting Equation (\ref{eq:AGNlight6}) into Equation (\ref{generaleq}), given $\hat{\eda}$, $\hat{\zda}$ and $\hat{\s}$, we obtain
\begin{equation}
\label{eq:AGNlight2}
\data^{\text{P}}={\burring}\hat{\sersic}+{\burring}\hat{\lensing}\hat{\s}+{\mapping}\PSF_{[0]},
\end{equation}
where $\hat{\lensing}={\lensing}(\hat{\zda})$, which is kept fixed at this step.  Note that the
$\burring$ matrix here is based on the initial PSF model, before the
multi-Gaussian fitting.  
The posterior of the PSF and AGN parameters is given by
\begin{equation}
\label{eq:AGNlight4}
{P(\xivec,\tauvec|\data,\hat{\eda},\hat{\zda})} = \frac{{P(\data|\hat{\eda},\hat{\zda},\xivec,\tauvec)}{P(\xivec,\tauvec)}}{P(\data|\hat{\eda},\hat{\zda})},
\end{equation}
where $\xivec$ represents the parameters of the Gaussian profiles in 
Equation (\ref{eq:AGNlight3}) that yield $\PSF_{[0]}$; $\tauvec$ 
are the amplitudes and the positions of the AGN, which are coded in $\mapping$.
The likelihood of Equation (\ref{eq:AGNlight4}) is 
\begin{equation}
\label{eq:AGNlight5}
{P(\data|\hat{\eda},\hat{\zda},\xivec,\tauvec)} = \frac{\mathrm{exp}[-E_{\text{D}}(\data|\hat{\eda},\hat{\zda},\xivec,\tauvec)]}{Z_{\text{D}}},
\end{equation}
where
\begin{displaymath}
E_{\text{D}}(\data|\hat{\eda},\hat{\zda},\xivec,\tauvec)\qquad\qquad\qquad\qquad\qquad\qquad
\end{displaymath}

\begin{displaymath}
=\frac{1}{2}({\data-{\burring}\hat{\sersic}-{\burring}\hat{\lensing}\hat{\s}-{\mapping}\PSF_{[0]}})^{\text{T}}\qquad
\end{displaymath}

\begin{equation}
\label{eq:AGNlight7}
\qquad\cdot\,{\bf C}_{\text{D}}^{-1}({\data-{\burring}\hat{\sersic}-{\burring}\hat{\lensing}\hat{\s}-{\mapping}\PSF_{[0]}}),
\end{equation}
and $Z_{\text{D}}=(2\pi)^{N_{\text{d}}/2}($det ${{\bf C}_{\text{D}}})^{1/2}$.
We denote $\hat{\xivec}$ and $\hat{\tauvec}$ as the maximum likelihood
parameters (that maximizes Equation
(\ref{eq:AGNlight4})) from which we can obtain the optimal AGN light on 
the image, given the optimized source and lens mass models from the previous sections.

\subsection{Pixelated Fine Structure of AGN light}
\label{sec:method:iterative}

In this section, we introduce the inner loop which aims at extracting
the fine structure, $\delta\PSF_{[0]}$, in Equation (\ref{eq:AGNlight1}) 
by using a correction grid. We show it visually in Figure 
\ref{fig:pixgrid_overlap}.
The goal of the inner loop is to incorporate most of the fine structures 
into the PSF model; then in the outer loop, we can use the updated PSF model
obtained from the inner loop to remodel all the light components (which
require a given PSF model).
Since this section is the starting point of the inner loop and outer loop, 
the $\hat{\eda}$, $\hat{\zda}$, $\hat{\s}$, $\hat{\xivec}$, and $\hat{\tauvec}$ 
we get by optimizing Equations (\ref{eq:lenslight1}), (\ref{eq:arclight1}), 
and (\ref{eq:AGNlight4}) are actually the zero-th outer loop iteration and 
the zero-th inner loop iteration, which we denote by $\hat{\eda}_{[0]}$, 
$\hat{\zda}_{[0]}$, $\hat{\s}_{[0]}$, $\hat{\xivec}_{[0,0]}$, and $\hat{\tauvec}_{[0,0]}$.

\subsubsection{PSF Correction for Each Iteration \\\ \qquad\qquad\qquad(Inner Loop: Step 3)}
\label{sec:method:iterative:eachiteration}
In general, given $\hat{\eda}_{[\ol]}$, $\hat{\zda}_{[\ol]}$, $\hat{\s}_{[\ol]}$, 
$\hat{\xivec}_{[\ol,\il]}$\footnote{$\hat{\xivec}_{[\ol,\il]}$ is only present 
when $\ol=\il=0$, which corresponds to parameters of the Gaussian profiles in 
Equation (\ref{eq:AGNlight3})},
and $\hat{\tauvec}_{[\ol,\il]}$, where $\il$ is the iteration number of 
the inner loop and $\ol$ is the iteration number of the outer loop,
we can write out the equation as
\begin{displaymath}
\data^{\text{P}}={{\burring}_{[\ol]}\hat{\sersic}_{[\ol]}+{\burring}_{[\ol]}\hat{\lensing}_{[\ol]}\hat{\s}_{[\ol]}}\qquad\qquad\qquad\qquad\qquad\qquad\quad
\end{displaymath}
\begin{equation}
\label{eq:fine_structure0}
+\hat{\mapping}_{[\ol,\il]}(\hat{\PSF}_{[\ol,\il]}+\Tvec_{[\ol,\il]}\delta\PSF_{[\ol,\il]})  \equiv \data^{\text{P}}_{~\text{correction}},
\end{equation}
where 
\begin{displaymath}
\hat{\PSF}_{[\ol,\il]} = \left\{ \begin{array}{ll} 
  \PSF_{[\ol,\il]}(\hat{\xivec}_{[\ol,\il]}) & \textrm{if $\ol=\il=0$} \\
  \PSF_{[\ol,\il]} & \textrm{otherwise}
 \end{array} \right.
\end{displaymath}
$\hat{\sersic}_{[\ol]}={\sersic}(\hat{\eda}_{[\ol]})$, 
$\hat{\lensing}_{[\ol]}={\lensing}(\hat{\zda}_{[\ol]})$, 
$\hat{\mapping}_{[\ol,\il]}={\mapping}(\hat{\tauvec}_{[\ol,\il]})$,
${\burring}_{[\ol]}$ is the $\ol^{\rm th}$ blurring matrix 
(we explain how to get the $\ol^{\rm th}$ blurring matrix in Section 
\ref{sec:method:Outer_loop}), $\Tvec_{[\ol,\il]}$ is a matrix which makes 
$\delta\PSF_{[\ol,\il]}$ the same length as $\hat{\PSF}_{[\ol,\il]}$ by 
padding with zeros 
(we show it visually in Appendix \ref{app:b}), and
$\delta\PSF_{[\ol,\il]}$ is the fine  
structure we want to obtain by the end of this section.
 
The posterior of $\delta\PSF_{[\ol,\il]}$ is
\begin{displaymath}
P({\delta\PSF_{[\ol,\il]}}|\data,\hat{\eda}_{[\ol]}, \hat{\zda}_{[\ol]}, \hat{\xivec}_{[\ol,\il]},\hat{\tauvec}_{[\ol,\il]},\lambda_{\delta\PSFsca,[\ol,\il]},R) \qquad\qquad\qquad
\end{displaymath}
\begin{displaymath}
= \frac{{P(\data|\delta\PSF_{[\ol,\il]},\hat{\eda}_{[\ol]}, \hat{\zda}_{[\ol]}, \hat{\xivec}_{[\ol,\il]},\hat{\tauvec}_{[\ol,\il]})}}{P(\data|\lambda_{\delta\PSFsca,[\ol,\il]},\hat{\eda}_{[\ol]}, \hat{\zda}_{[\ol]}, \hat{\xivec}_{[\ol,\il]},\hat{\tauvec}_{[\ol,\il]},R)}\qquad
\end{displaymath}
\begin{equation}
\label{eq:fine_structure1}
\cdot P(\delta\PSF_{[\ol,\il]}|\lambda_{\delta\PSFsca,[\ol,\il]},R),\qquad\qquad
\end{equation}
where $P(\delta\PSF_{[\ol,\il]}|\lambda_{\delta\PSFsca,[\ol,\il]},R)$ is the prior on 
$\delta\PSF_{[\ol,\il]}$ given 
{\{$\lambda_{\delta\PSFsca,[\ol,\il]},R$\}} with $R$ denoting a
particular form of ``regularization'' on $\delta\PSF_{[\ol,\il]}$ and
$\lambda_{\delta\PSFsca,[\ol,\il]}$ characterising the strength of
the regularization. 
We can write the likelihood in Equation (\ref{eq:fine_structure1}) as 
\begin{displaymath}
P(\data|\delta\PSF_{[\ol,\il]},\hat{\eda}_{[\ol]}, \hat{\zda}_{[\ol]}, \hat{\xivec}_{[\ol,\il]},\hat{\tauvec}_{[\ol,\il]})\qquad\qquad\qquad\qquad
\end{displaymath}
\begin{equation}
\label{eq:fine_structure2}
= \frac{\mathrm{exp}[-E_{\text{D},\text{mAc},[\ol,\il]}(\data|\delta\PSF_{[\ol,\il},\hat{\eda}_{[\ol]}, \hat{\zda}_{[\ol]}, \hat{\xivec}_{[\ol,\il]},\hat{\tauvec}_{[\ol,\il]})]}{Z_{\text{D,mAc}}},
\end{equation}
where ``mAc'' stands for maskAGNcenter,
\begin{displaymath}
{E_{\text{D},\text{mAc},[\ol,\il]}(\data|\delta\PSF_{[\ol,\il]},\hat{\eda}_{[\ol]}, \hat{\zda}_{[\ol]}, \hat{\xivec}_{[\ol,\il]},\hat{\tauvec}_{[\ol,\il]})} \qquad\qquad
\end{displaymath}

\begin{equation}
\label{eq:fine_structure3}
=\frac{1}{2}({\data-\data^{\text{P}}_{~\text{correction}}})^{\text{T}}{\bf C}_{\text{D,mAc}}^{-1}({\data- \data^{\text{P}}_{~\text{correction}}}),
\end{equation}
and $Z_{\text{D,mAc}}=(2\pi)^{N_{\rm d}/2}(\text{det}\,{{\bf C}_{\text{D,mAc}}})^{1/2}$ is the normalization for the probability. We discuss the mAc (maskAGNcenter)
regions in Appendix \ref{app:a} 

The prior/regularization we impose in Equation
(\ref{eq:fine_structure1}) on the correction grid (fine structure of
PSF) is to prevent the correction grid from absorbing the noise in the
observed image. 
We express the prior in the following form
\begin{displaymath}
P({\delta\PSF_{[\ol,\il]}}|\lambda_{\delta\PSFsca,[\ol,\il]},R) \qquad\qquad\qquad\qquad
\end{displaymath}
\begin{equation}
\label{eq:fine_structure4}
= \frac{\mathrm{exp}(-\lambda_{\delta\PSFsca,[\il]} E_{\delta\PSFsca,[\ol,\il]}({\delta\PSF_{[\ol,\il]}}|{R}))}{Z_{\delta\PSFsca,[\ol,\il]}(\lambda_{\delta\PSFsca,[\ol,\il]})},
\end{equation}
where $\lambda_{\delta\PSFsca,[\ol,\il]}$ is the regularization constant
of correction,  
$Z_{\delta\PSFsca,[\ol,\il]}(\lambda_{\delta\PSFsca,[\ol,\il]})=
\int d^{N_{\delta\PSFsca,[\ol,\il]}}{\delta\PSF_{[\ol,\il]}}$ exp
$(-\lambda_{\delta\PSFsca,[\ol,\il]}E_{\delta\PSFsca,[\ol,\il]})$ is the normalization 
of the prior probability distribution (note that the optimal 
$\lambda_{\delta\PSFsca,[\ol,\il]}$ is not determined yet), and
$N_{\delta\PSFsca,[\ol,\il]}$ 
is the number of pixels of the correction grid. We use the curvature form 
for the function $E_{\delta\PSFsca,[\ol,\il]}$, which is discussed in 
\citet{SuyuEtal06}. 

Again, it is easy to understand that we want to maximize Equation (\ref{eq:fine_structure1}). We obtain the most probable solution
\begin{equation}
\label{eq:fine_structure5}
\delta\PSF_{[\ol,\il]}=({\bf
  F}+{\lambda_{\delta\PSFsca,[\ol,\il]}}{\bf H})^{-1} ({\mapping}_{[\ol,\il]}{\Tvec}_{[\ol,\il]})^{\text{T}}{\bf C}_{\text{D,mAc}}^{-1}{\boldsymbol{u}},
\end{equation}
where
\begin{displaymath}
{\bf F}=\nabla\nabla E_{\text{D,mAc},[\ol,\il]}\qquad\qquad\qquad\qquad\quad
\end{displaymath} 
\begin{equation}
\label{eq:fine_structure6}
\qquad\qquad\quad={\Tvec}_{[\ol,\il]}^{\text{T}}({\mapping}_{[\ol,\il]}^{\text{T}}{\bf C}_{\text{D,mAc}}^{-1}~{\mapping}_{[\ol,\il]}){\Tvec}_{[\ol,\il]},
\end{equation}

\begin{equation}
\label{eq:fine_structure7}
{\bf H}=\nabla{\nabla{E_{\delta\PSFsca,[\ol,\il]}}},\quad\,\,\,\,
\end{equation}

\begin{equation}
\label{eq:fine_structure8}
{\boldsymbol{u}}=\data-{\burring}_{[\ol]}\hat{\sersic}_{[\ol]}-{\burring}_{[\ol]}\hat{\lensing}_{[\ol]}\s_{[\ol]}-\hat{\mapping}_{[\ol,\il]}\hat{\PSF}_{[\ol,\il]},
\end{equation}
and
\begin{equation}
\label{eq:fine_structure9}
\nabla\equiv\frac{\partial}{\partial\delta\PSF_{[\ol,\il]}}.\qquad\quad\,\,
\end{equation}

Now, we go back to find the optimal regularization constant; that is, 
we want to maximize
\begin{equation}
\label{eq:fine_structure10}
P({\lambda_{\delta\PSFsca,[\ol,\il]}|\data,{R}})=\frac{P({\data|{R},\lambda_{\delta\PSFsca[,[\ol,\il]}})P({\lambda_{\delta\PSFsca,[\ol,\il]}})}{P(\data|{R})}
\end{equation}
using Bayes' rule. If we assume a flat prior in 
log~$\lambda_{\delta\PSFsca,[\ol,\il]}$, we want to maximize 
$P({\data|{R},\lambda_{\delta\PSFsca,[\ol,\il]}})$, which is the 
evidence in equation (\ref{eq:fine_structure1}). Following the results 
from \citet{SuyuEtal06}, we get
\begin{displaymath}
2\hat\lambda_{\delta\PSFsca,[\ol,\il]}{E_{\delta\PSFsca,[\ol,\il]}}(\delta\PSF_{[\ol,\il]})\qquad\qquad\qquad\qquad\qquad\qquad
\end{displaymath} 
\begin{equation}
\label{eq:fine_structure11}
=N_{\delta\PSFsca,[\ol,\il]}-\hat\lambda_{\delta\PSFsca,[\ol,\il]}\mathrm{Tr}[{{({\bf F}+{\hat\lambda_{\delta\PSFsca,[\ol,\il]}}{\bf H})^{-1} {\bf H}}}],
\end{equation}
where Tr denotes the trace and $\hat\lambda_{\delta\PSFsca,[\ol,\il]}$ is the 
optimal 
regularization constant. If we set $\il=0$ (zeroth iteration of the fine 
structure), we obtain $\delta\PSF_{[\ol,0]}$. Due to the sharp intensity of 
the AGN center, the residuals there are much stronger than the peripheral area.
If we directly extract 
the full correction grid,
the regularization intends 
to under-regularize on the peripheral area and over-regularize on the center. 
To avoid this problem, at first, we extract the correction only around the AGN 
center; that is, we start from small $N_{\delta\PSFsca,[\ol,\il]}$ (half light radius or smaller) and increase it 
gradually (around 1.2 times previous size each time) as we obtain $\delta\PSF_{[\ol,\il]}$. We show the idea 
in Figure \ref{fig:pixgrid_overlap} 
(note that the indices on $\delta\PSF$ in the figure are labeling the pixels, 
rather than the iteration numbers).

Since every iteration of $\delta\PSF_{[\ol,\il]}$ has their own fine structure 
(correction) uncertainty, according to \citet{SuyuEtal06}, we also take as 
estimates of the $1\sigma$ uncertainty on each pixel value the square root 
of the corresponding diagonal element of the covariance matrix given by
\begin{equation}
\label{eq:fine_structure12}
{\bf C}_{\delta\PSFsca,[\ol,\il]}=({\bf F}+{\hat\lambda_{\delta\PSFsca,[\ol,\il]}}{\bf H})^{-1}.
\end{equation}

\subsubsection{Add Fine Structure into Global Structure \\ \quad\qquad\qquad\qquad(Inner Loop: Step 4)}
\label{sec:method:add}
We start with the zeroth inner loop iteration, 
by setting $\il=0$, of the global structure, $\PSF_{[\ol,0]}$, and fine structure,
 $\delta\PSF_{[\ol,0]}$ (which we can obtain by following the previous two
 sections). We then add the fine structure into the global structure 
by defining
\begin{equation}
\label{eq:add1}
\PSF_{[\ol,1]}=\PSF_{[\ol,0]}+\Tvec_{[\ol,0]}\delta\PSF_{[\ol,0]},
\end{equation}
where $\PSF_{[\ol,1]}$ is the first iteration in inner loop. More generally, 
we define the $\il+1^{\rm th}$ iteration of the PSF as
\begin{equation}
\label{eq:add2}
\PSF_{[\ol,\il+1]}=\PSF_{[\ol,\il]}+\Tvec_{[\ol,\il]}\delta\PSF_{[\ol,\il]}.
\end{equation}

We recalculate the
AGN parameters every time after getting a new 
$\PSF_{[\ol,\il+1]}$, so given the same $\hat{\eda}_{[\ol]}$ and $\hat{\zda}_{[\ol]}$ 
in Equation (\ref{eq:fine_structure0}), the posterior of 
the AGN parameters is given by
\begin{displaymath}
{P(\tauvec_{[\ol,\il+1]}|\data,\hat{\eda}_{[\ol]},\hat{\zda}_{[\ol]})}\qquad\qquad\qquad\qquad
\end{displaymath}
\begin{equation}
\label{eq:add3}
=\frac{{P(\data|\hat{\eda}_{[\ol]},\hat{\zda}_{[\ol]},\tauvec_{[\ol,\il+1]})}{P(\tauvec_{[\ol,\il+1]})}}{P(\data|\hat{\eda}_{[\ol]},\hat{\zda}_{[\ol]})}.
\end{equation}
(recall that $\tauvec_{[\ol,\il+1]}$ represents the relative amplitudes and 
the positions of the AGNs in the $\ol^{\rm th}$ outer loop iteration, and 
$\il+1^{\rm th }$ inner loop iteration). The likelihood of Equation 
(\ref{eq:add3}) is 

\begin{displaymath}
{P(\data|\hat{\eda}_{[\ol]},\hat{\zda}_{[\ol]},\tauvec_{[\ol,\il+1]})}\qquad\qquad\qquad\qquad\qquad\qquad
\end{displaymath}

\begin{equation}
\label{eq:add4}
=\frac{\mathrm{exp}[-E_{\text{D},[\ol,\il+1]}(\data|\hat{\eda}_{[\ol]},\hat{\zda}_{[\ol]},\tauvec_{[\ol,\il+1]})]}{Z_{\text{D}}},
\end{equation}
where
\begin{displaymath}
E_{{\rm D},[\ol,\il+1]}(\data|\hat{\eda}_{[\ol]},\hat{\zda}_{[\ol]},\tauvec_{[\ol,\il+1]})
\end{displaymath}
\begin{equation}
\label{eq:add5}
=\frac{1}{2}({\data-\boldsymbol{\Omega}})^{\text{T}}{\bf C}_{\text{D}}^{-1}({\data-\boldsymbol{\Omega}})
\end{equation}
with
\begin{equation}
\boldsymbol{\Omega}={{\burring}_{[\ol]}\hat{\sersic}_{[\ol]}}+{{\burring}_{[\ol]}\hat{\lensing}_{[\ol]}\hat{\s}_{[\ol]}}+{\mapping}_{[\ol,\il+1]}{\PSF}_{[\ol,\il+1]},
\end{equation}
and $Z_{\text{D}}$, as usual, is $(2\pi)^{N_{\rm d}/2}($det ${{\bf C}_{\text{D}}})^{1/2}$.
After maximizing Equation (\ref{eq:add3}), we obtain $\hat{\tauvec}_{[\ol,\il+1]}$. We then replace the $\hat{\tauvec}_{[\ol,\il]}$ from the previous iteration with 
the $\hat{\tauvec}_{[\ol,\il+1]}$ we just obtained, and conduct  
the next inner loop iteration.

\subsubsection{The Criteria to Stop the Inner Loop.}
\label{sec:method:innerloopstop}

 During every inner loop, we gradually increase the size, 
$N_{\delta\PSFsca,[\ol,\il]}$, of the correction grid. Then, if 
(1) there is no residuals outside the correction grid, 
(2) Equation (\ref{eq:fine_structure5}) has no intensity, and 
(3) Equation (\ref{eq:add5}) no longer decreases, we stop the inner loop. 
Assuming we have $N_{\text{inner}}$ iterations in the inner loop, we obtain 
$\PSF_{[\ol,N_{\text{inner}}]}$ and
$\hat{\tauvec}_{[\ol,N_{\text{inner}}]}$. We then define
\begin{equation}
\label{eq:add6}
\PSF_{[\ol,N_{\text{inner}}]}\equiv\PSF_{[\ol+1,0]}\equiv\PSF_{[\ol+1]}
\end{equation}
and
\begin{equation}
\label{eq:add7}
\hat{\tauvec}_{[\ol,N_{\text{inner}}]}\equiv\hat{\tauvec}_{[\ol+1,0]}\equiv\hat{\tauvec}_{[\ol+1]}.
\end{equation}

\subsection{Lens Modeling with updated PSF}
\label{sec:method:lensModelUpdated}
The goal of the outer loop in Figure \ref{fig:diagram} is to remodel
all the light components with the updated PSF; that is, we want to obtain 
a better lens light model and arc light model with the new blurring matrix, 
and the underlying fine structure can then be revealed. 

\subsubsection{Update the Blurring Matrix and the Image Covariance Matrix (Outer Loop: Step 6)}
\label{sec:method:Outer_loop}
{\it Blurring matrix}~: After obtaining the last version of the PSF from 
Section \ref{sec:method:innerloopstop}, we update the blurring matrix, 
${\burring}$, in Equation (\ref{generaleq}). In order to accelerate modeling 
speed, which highly depends on the size of the PSF for convolution 
of the extended images, 
we choose the central $l_{[\ol]} \times l_{[\ol]}$ pixels of the
updated PSF grid
(that has $N_{\delta\PSFsca,[\ol,N\text{inner}]}$ pixels) as the new
PSF to construct ${\burring}_{[\ol+1]}$ for the spatially extended images.\footnote{For example, we choose $l_{[\ol]}=n l_{[1]}$}

{\it Image covariance matrix}~: We accumulate the uncertainty of the PSF pixel grid from 
every inner loop. The accumulated uncertainty is 
\begin{equation}
\label{eq:Outer_loop1}
n^{2}_{\delta\PSFsca,[\ol+1],k}=\sum_{\il=0}^{N_{\rm inner}}\sum_{i}{\Tsca}_{[\ol,\il],ki}{\rm C}_{\delta\PSFsca,[\ol,\il],ij}\delta_{ij},
\end{equation}
where ${\Tsca}_{[\ol,\il],ki}$ is the element at $k$ row and $i$ column of 
${\Tvec}_{[\ol,\il]}$, ${\rm C}_{\delta\PSFsca,[\ol,\il],ij}$ is the element at $i$ 
row and $j$ column of ${\bf C}_{\delta\PSFsca,[\ol,\il]}$, and $\delta_{ij}$ is 
the Kronecker delta.
The element of the $\ol+1^{\rm th}$ noise vector is defined as
\begin{equation}
\label{noise_iteration}
n_{[\ol+1],\mu}=\sqrt{n_{\mu}^{2}+\sum_{k}\hat{\rm M}_{[\ol+1],\mu k}n_{\delta\PSFsca,[\ol+1],k}^{2}},
\end{equation} 
which is characterized by the covariance 
matrix ${\bf C}_{\text{D},[\ol+1]}$\footnote{The purpose of updating the
  the image covariance matrix is to speed up the modeling to the final
  answer since the correction uncertainty that we add into the image covariance
  matrix is around AGN; that is, we weight the arc region more. However,
  in the end, if there is no ``correction'', Equation
  (\ref{eq:Outer_loop1}) is close to zero.}. Note that $n_{\mu}$ is the element 
of the original data noise vector.

\subsubsection{Lens Modeling with All Light Components \\ \qquad\qquad\qquad\qquad(Outer loop: Step 2)}
\label{sec:method:Outer_loop:modelall}
In general, when executing the next iteration of outer loop, we can express 
Equation (\ref{generaleq}) as
\begin{displaymath}
\label{eq:Outer_loop2.5}
\data^{\rm P}={{\burring}_{[\ol+1]}{\sersic}_{[\ol+1]}}+{{\burring}_{[\ol+1]}{\lensing}_{[\ol+1]}\s_{[\ol+1]}}+{\mapping}_{[\ol+1]}\PSF_{[\ol+1]}
\end{displaymath}
\begin{equation}
\equiv\data^{\text{P}}_{~\text{total}}.\qquad\qquad\qquad\qquad\qquad\qquad\qquad\qquad
\end{equation}
The posterior can be written as
\begin{displaymath}
P(\eda_{[\ol+1]},\zda_{[\ol+1]},{\tauvec_{[\ol+1]}}|\data,\Delta\tvec)\qquad\qquad\qquad\qquad\qquad\qquad\qquad\quad
\end{displaymath}
\begin{equation}
\label{eq:Outer_loop3}
\propto P(\data,\Delta\tvec|\eda_{[\ol+1]},\zda_{[\ol+1]},{\tauvec}_{[\ol+1]})P(\eda_{[\ol+1]},\zda_{[\ol+1]},{\tauvec}_{[\ol+1]}).
\end{equation}
The likelihood of the data can be expressed as
\begin{displaymath}
P({\data},{\Delta\tvec}|{\eda_{[\ol+1]},\zda_{[\ol+1]},{\tauvec}_{[\ol+1]}})\qquad\qquad\qquad\qquad\qquad\qquad\quad
\end{displaymath}
\begin{equation}
\label{eq:Outer_loop4}
= \int\rm{d}{\s_{[\ol+1]}}\:{P({\data},{\Delta\tvec}|{\eda}_{[\ol+1]},{\zda}_{[\ol+1]},{\tauvec}_{[\ol+1]},{\s_{[\ol+1]}})}{P(\s_{[\ol+1]})},
\end{equation}
where
\begin{displaymath}
{P(\data,\Delta\tvec|{\eda}_{[\ol+1]},{\zda}_{[\ol+1]},{\tauvec}_{[\ol+1]},\s_{[\ol+1]})} \qquad \qquad \qquad \qquad \qquad
\end{displaymath}

\begin{displaymath}
=\frac{\mathrm{exp}[-E_{\text{D},[\ol+1]}(\data|{\eda}_{[\ol+1]},{\zda}_{[\ol+1]},{\tauvec}_{[\ol+1]},\s_{\ol+1})]}{Z_{\text{D},[\ol+1]}}\qquad 
\end{displaymath}

\begin{displaymath}
\cdot\prod_{i=1}^{N_{\text{AGN}}}\frac{1}{\sqrt{2\pi}\sigma_{{\rm
      AGN},i}}\mathrm{exp}\left(-\frac{|\hat{\boldsymbol{\theta}}_{{\rm
        AGN},i,[\ol+1]}-\boldsymbol{\theta}_{\rm{AGN},i,[\ol+1]}^{\rm
      P}|^{2}}{2\sigma_{{\rm AGN},i}^{2}}\right)
\end{displaymath}
\begin{equation}
\label{eq:Outer_loop6}
\qquad \qquad
\cdot\prod_{i=1}\frac{1}{\sqrt{2\pi}\sigma_{\Delta\tsca,i}}\mathrm{exp}\left[-\frac{(\Delta\tsca_{i}-\Delta\tsca_{i,[\ol+1]}^{\rm
      P})^{2}}{2\sigma_{\Delta\tsca,i}^{2}}\right],
\end{equation}

\begin{displaymath}
{E_{\text{D},[\ol+1]}(\data|\eda_{[\ol+1]},\zda_{[\ol+1]},\tauvec_{[\ol+1]},\s_{[\ol+1]})}\qquad
\end{displaymath}

\begin{equation}
\label{eq:Outer_loop7}
=\frac{1}{2}({\data-\data^{\text{P}}_{~\text{total}}})^{\text{T}}{\bf C}_{\text{D},[\ol+1]}^{-1}({\data-\data^{\text{P}}_{~\text{total}}}),
\end{equation}
where

\begin{equation}
Z_{\text{D},[\ol+1]}=(2\pi)^{N_{\datasca}/2}(\text{det }{\bf C}_{\text{D},[\ol+1]})^{1/2}
\end{equation}
is the normalization for the probability, and
$\hat{\boldsymbol{\theta}}_{{\rm AGN},i,[\ol+1]}
=\boldsymbol{\theta}_{{\rm AGN},i}(\hat{\tauvec}_{[\ol+1]})$.

After maximizing Equation (\ref{eq:Outer_loop3}), we obtain 
$\hat{\eda}_{[\ol+1]}$, $\hat{\zda}_{[\ol+1]}$, and $\hat{\tauvec}_{[\ol+1]}$. 
Then, we replace the $\hat{\eda}_{[\ol]}$, $\hat{\zda}_{[\ol]}$, and 
$\hat{\tauvec}_{[\ol]}$ in Section \ref{sec:method:iterative} with 
$\hat{\eda}_{[\ol+1]}$, $\hat{\zda}_{[\ol+1]}$, and $\hat{\tauvec}_{[\ol+1]}$ 
and then execute the next set of inner loop iterations. 
If we have a total of $N$ outer loop iterations, 
we obtain the final ${\burring}_{[N]}$ and $\PSF_{[N]}$.

\begin{figure*}
\centering
\includegraphics[scale=0.5]{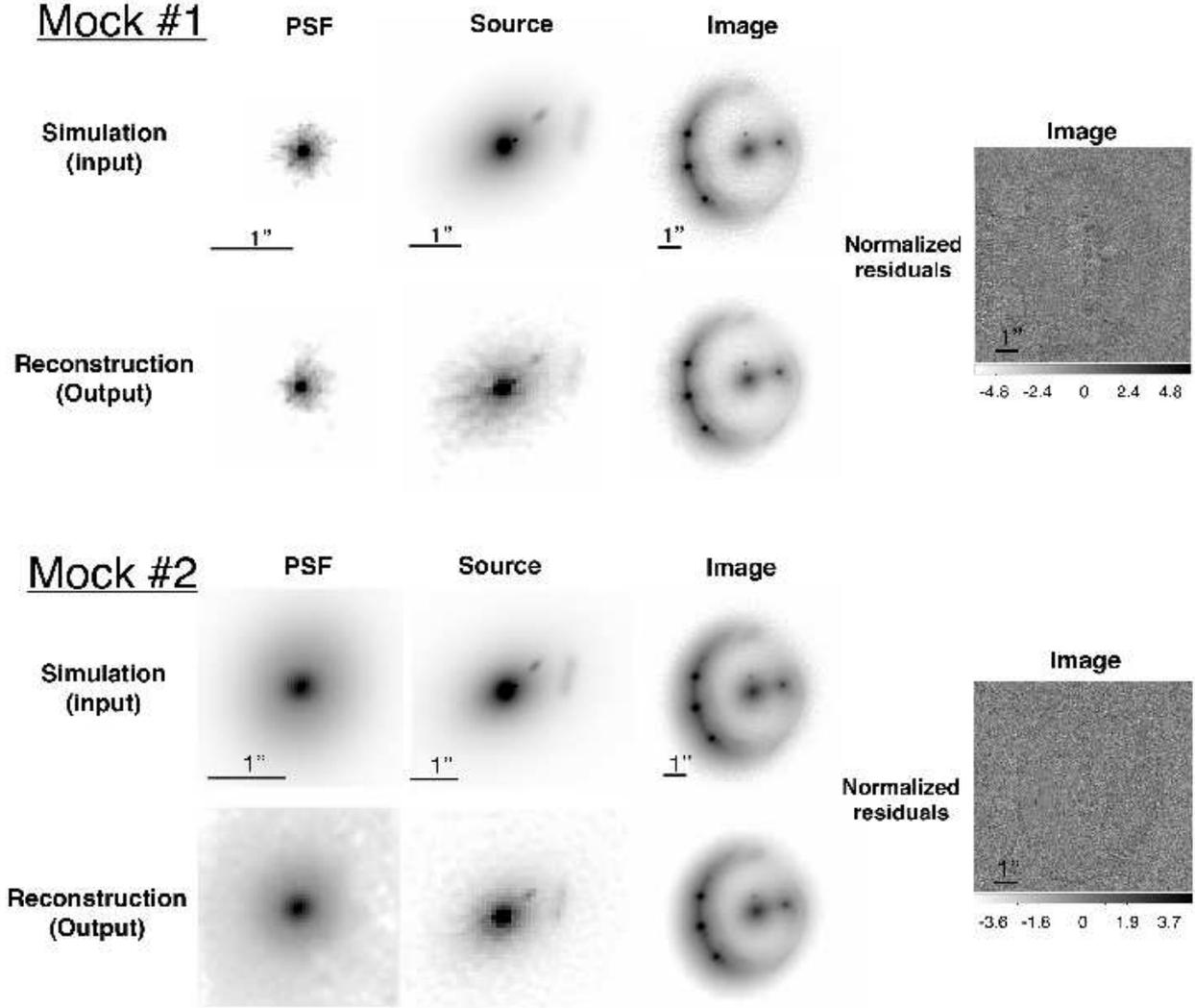}
\caption{The simulation (input), reconstruction (output), and 
  normalized residuals of mock \#1 and mock \#2. The left column shows the 
  input/output PSF, the middle left
  column shows the input/output sources (host galaxy of the AGN), the middle 
  right column shows
  the input/output images, and the right column shows the 
  normalized image residules (in units of the estimated pixel uncertainties). 
  Our PSF reconstruction method is
  able to reproduce both the global and fine structures in the PSF,
  yielding successful reconstructions of the background source 
  intensity and the lensed images. Both reduced $\chi^2$ are $\sim$ 1.
}
\label{fig:mock1_1}
\end{figure*}

\subsubsection{The Criteria to Stop the Outer Loop.}
\label{sec:method:outerloopstop}
We iterate the outer loop until Equation (\ref{eq:Outer_loop7})
does not decrease.\footnote{ $\frac{E_{\text{D},[\ol]}-E_{\text{D},[\ol+1]}}{E_{\text{D},[\ol]}}<0.2\%$} 
We also ensure that the size of the PSF ($l_{[\ol]} \times l_{[\ol]}$)
for convolution of the lens light and arc light is big enough.  Since
the AO PSF can have substantial wings that contribute significantly,
the size of the PSF in AO image is usually substantially larger than
those of \hst\ images.  We set the size of the PSF ($l_{[\ol]}\times
l_{[\ol]}$) such that the modeling results remain stable beyond this
PSF size.

\begin{figure*}
\centering
\includegraphics[scale=0.55]{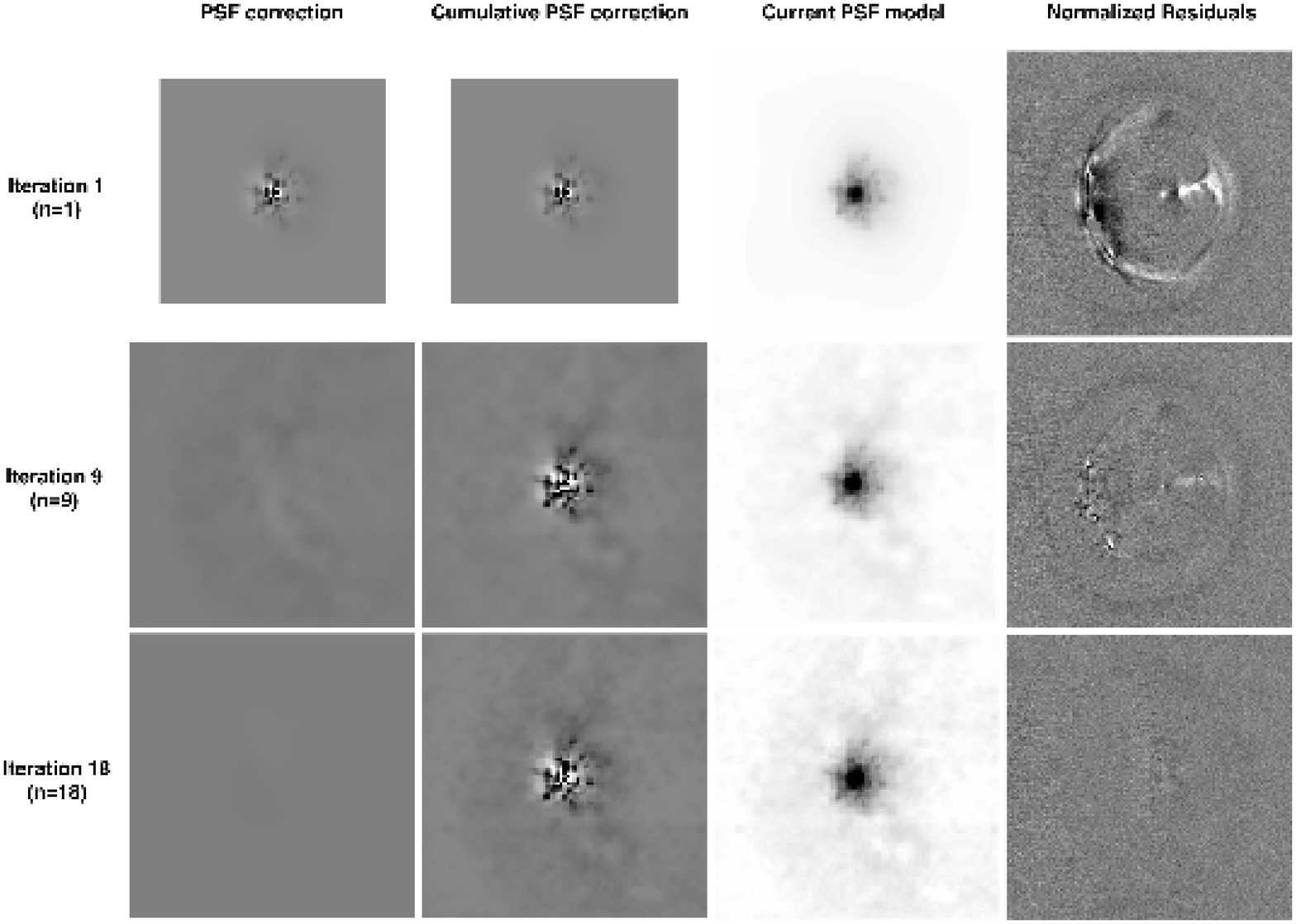}
\caption{We demonstrate the iterative reconstruction process. From the
  left to the right, we show the PSF correction, cumulative PSF
  correction, current PSF model, and normalized residuals after using
  the current PSF model at iteration $1$, $9$, and
  $18$. Since we sequentially 
  increase the PSF correction grid as we iterate, the size of the PSF
  correction grid at iteration $1$ is smaller than that of other
  iterations.} 
\label{fig:mock1_2}
\end{figure*}


\section{Demonstration and blind test}
\label{sec:blindtest}
In this section, we demonstrate the method using two mock data sets that 
are created with different PSFs, and show that we can recover the input 
parameters in both mocks by using the strategy in Section 
\ref{sec:method} together with {\sc Glee}. 
S.H.S.~simulates AO images that mimic the strong lensing system, \rxj, 
with two foreground lens galaxies and a background source comprised of
an AGN and its host galaxy. 
S.H.S.~ uses an elliptically symmetric power-law profile
to describe the main lens mass distribution and a pseudoisothermal
elliptic mass profile to describe the mass distribution of the
satellite galaxy. The background host
galaxy of the AGN is described by a S$\acute{\text{e}}$rsic profile 
with additional star-forming regions superposed, and the lens light 
distribution is based on a composite of two S$\acute{\text{e}}$rsic 
light profiles.
The simulated lensed images and background sources are shown in the
third and second column, respectively, of 
the first (mock \#1) and third (mock \#2) rows of Figure \ref{fig:mock1_1}.
The difference between the two mocks is their PSFs.  In mock \#1, the PSF is
taken to be a star observed with Keck's laser guide star adaptive
optics system (LGSAO) that is
relatively sharp and with a lot of structures (FWHM is $\sim0.03''$).  
In mock \#2, the PSF is relatively diffuse 
and without structures, which is similar to the PSF in the real data 
(FWHM is $\sim0.045''$). We show them in the first column of the first and
third rows of Figure \ref{fig:mock1_1}.  G.C.F.C.~does a blind test
of the PSF reconstruction method on 
mock \#1; that is, G.C.F.C.~does not know the input value at the
beginning, and S.H.S. only reveals the input value when G.C.F.C.~has completed
the analysis of mock \#1. Since the input value is the same in mock \#2,
G.C.F.C.~models mock \#2 by using the same strategy although the mock \#2 test
is performed after mock \#1 and therefore is not blinded.

\begin{figure*}
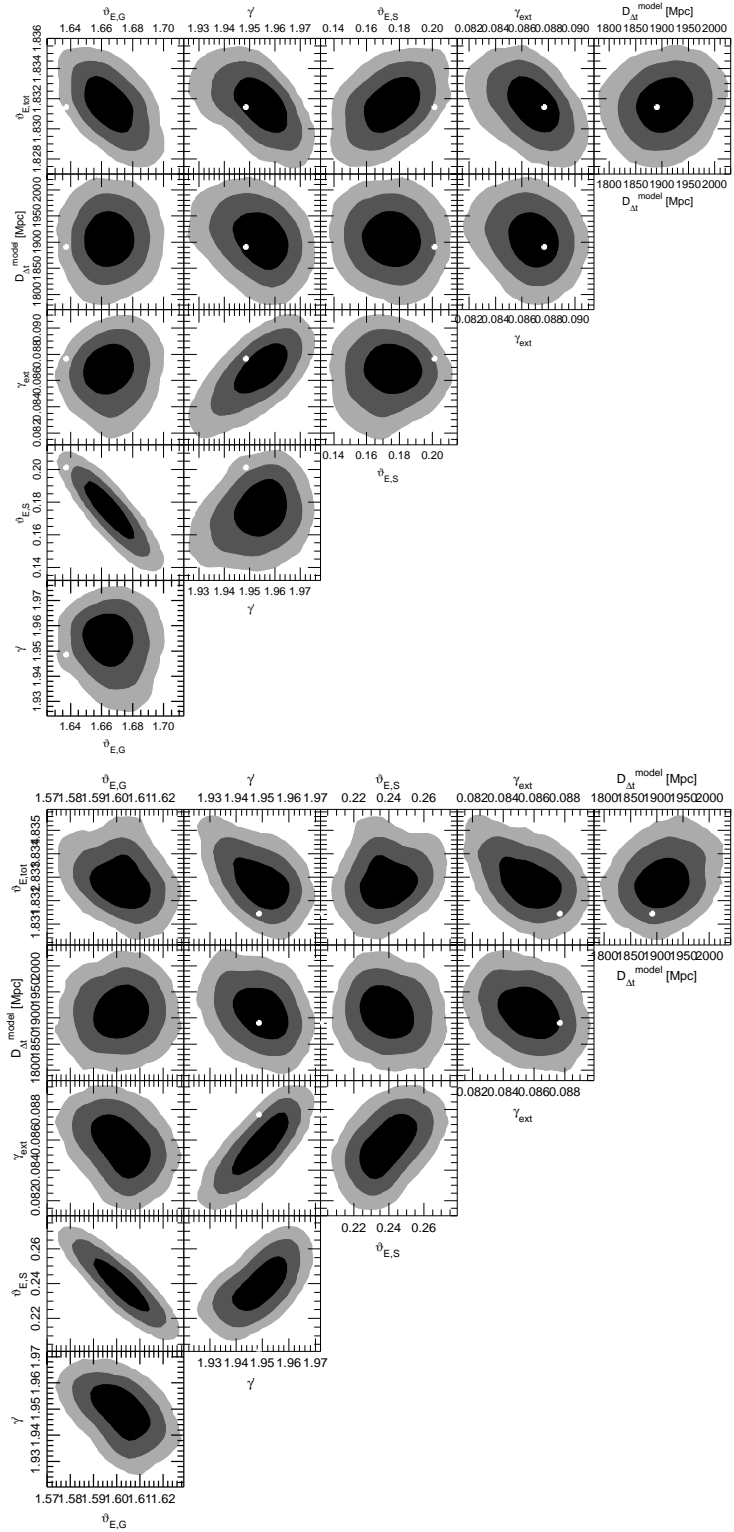

  \centering
  \includegraphics[width=0.55\textwidth, clip]{mock1_18_52_54_56_58_60_62_tot_2s+1_stats.eps}
  \includegraphics[width=0.55\textwidth, clip]{spemd1_sh_2sersic_8AGN_3_p59_s60_61_62_63_2s+1_stats.eps}
  \caption{ {\bf Upper panel}: the posterior probability distribution of the 
    key lens model parameters for mock \#1. We use the PSF size, 
    $19\times19$, for convolution of the spatially extended images of 
    the AGN host galaxy. We combine the different source resolutions: 
    $52\times52$, $54\times54$, $56\times56$, $58\times58$, $60\times60$, 
    and $62\times62$, and weight each chain equally. The contours/shades 
    mark the $68.3\%$, $95.4\%$, and $99.7\%$ credible regions. The white 
    dots are the input values. {\bf Lower panel}: the posterior probability 
    distribution of the key lens model parameters for mock \#2. We use the 
    PSF size, $59\times59$, for convolution of the spatially extended images 
    of the AGN host galaxy. We combine the different source resolutions: 
    $59\times59$, $60\times60$, $61\times61$, $62\times62$, and $63\times63$ 
    and weight each chain equally. We can recover the key lens parameters for 
    cosmography such as the modeled time-delay distance, total Einstein radius, 
    and external shear, despite the strong degeneracy between the Einstein 
    radii of the main and satellite galaxies (which consequently we do not 
    recover in mock \#2).
}
  \label{fig:mock1_sample}
\end{figure*}

\begin{figure*}
\centering
\includegraphics*[scale=0.65]{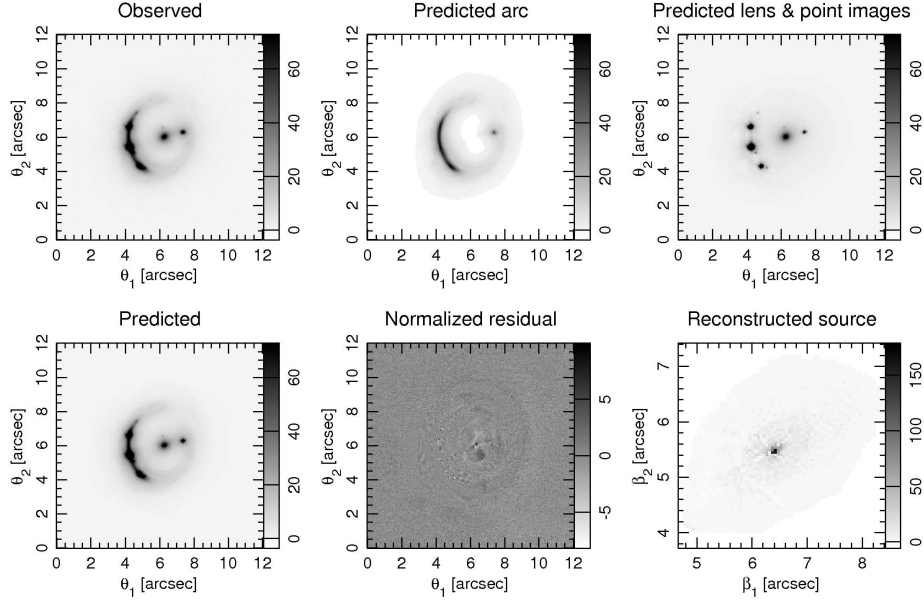}
\caption{\rxj\ AO image reconstruction of the most probable model with
  a source grid of $79\times79$ pixels and $69\times69$ PSF for
  convolution of spatially extended images. Top left: \rxj\ AO
  image. Top middle: predicted lensed image of the background AGN host
  galaxy. Top right: predicted light of the lensed AGNs, the bright compact 
  region: lensed images of a bright compact region in the AGN host galaxy,
  and the lens
  galaxies. Bottom left: predicted image from all components, which is
  a sum of the top-middle and top-right panels. Bottom middle: image
  residual, normalized by the estimated 1-$\sigma$ uncertainty of each
  pixel. Bottom right: the reconstructed host galaxy of the AGN in the
  source plane} 
\label{fig:real_demo}
\end{figure*}

\begin{figure*}
  \centering
  \includegraphics[width=0.3\textwidth, clip]{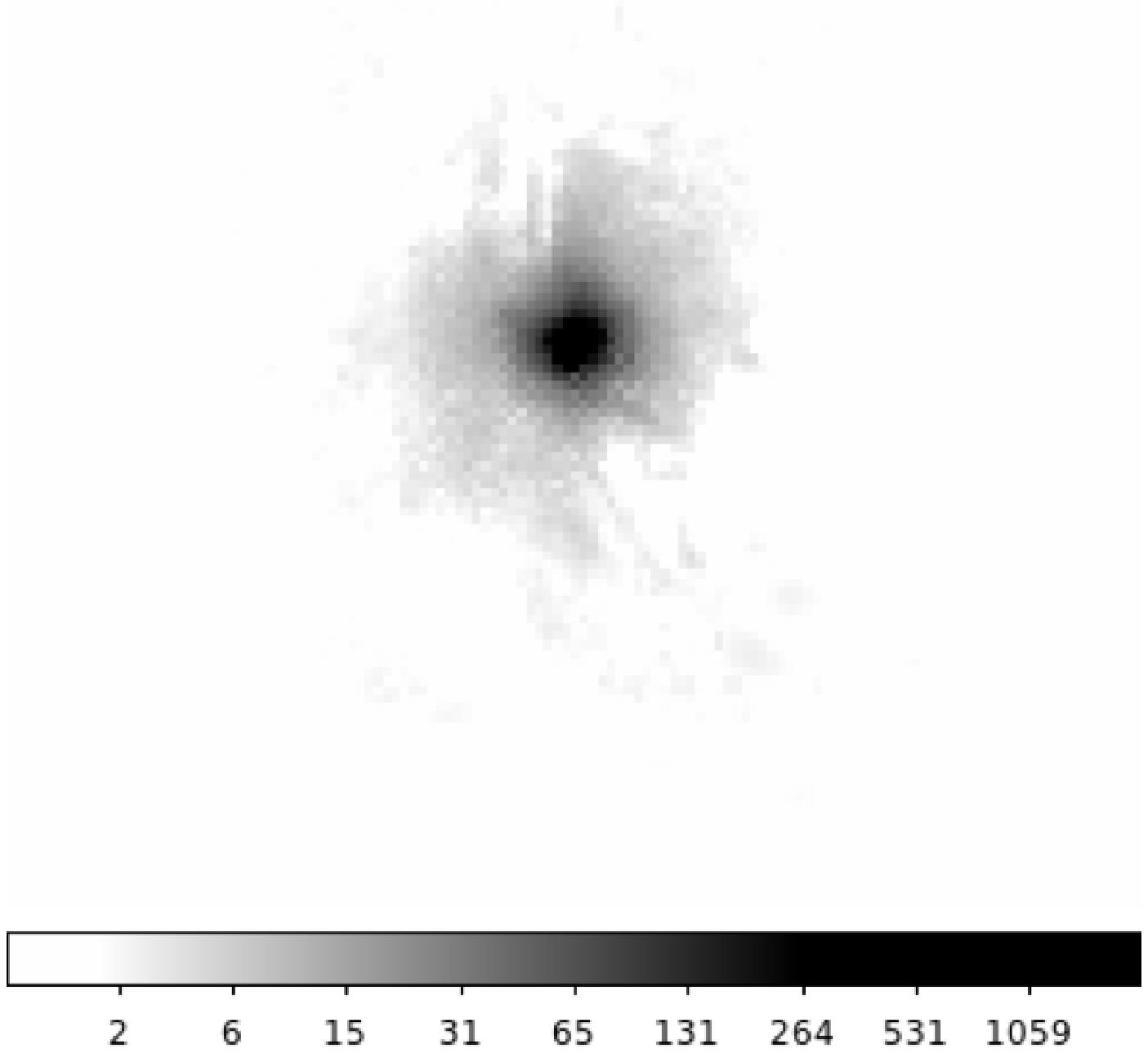}
  \includegraphics[width=0.4\textwidth, clip]{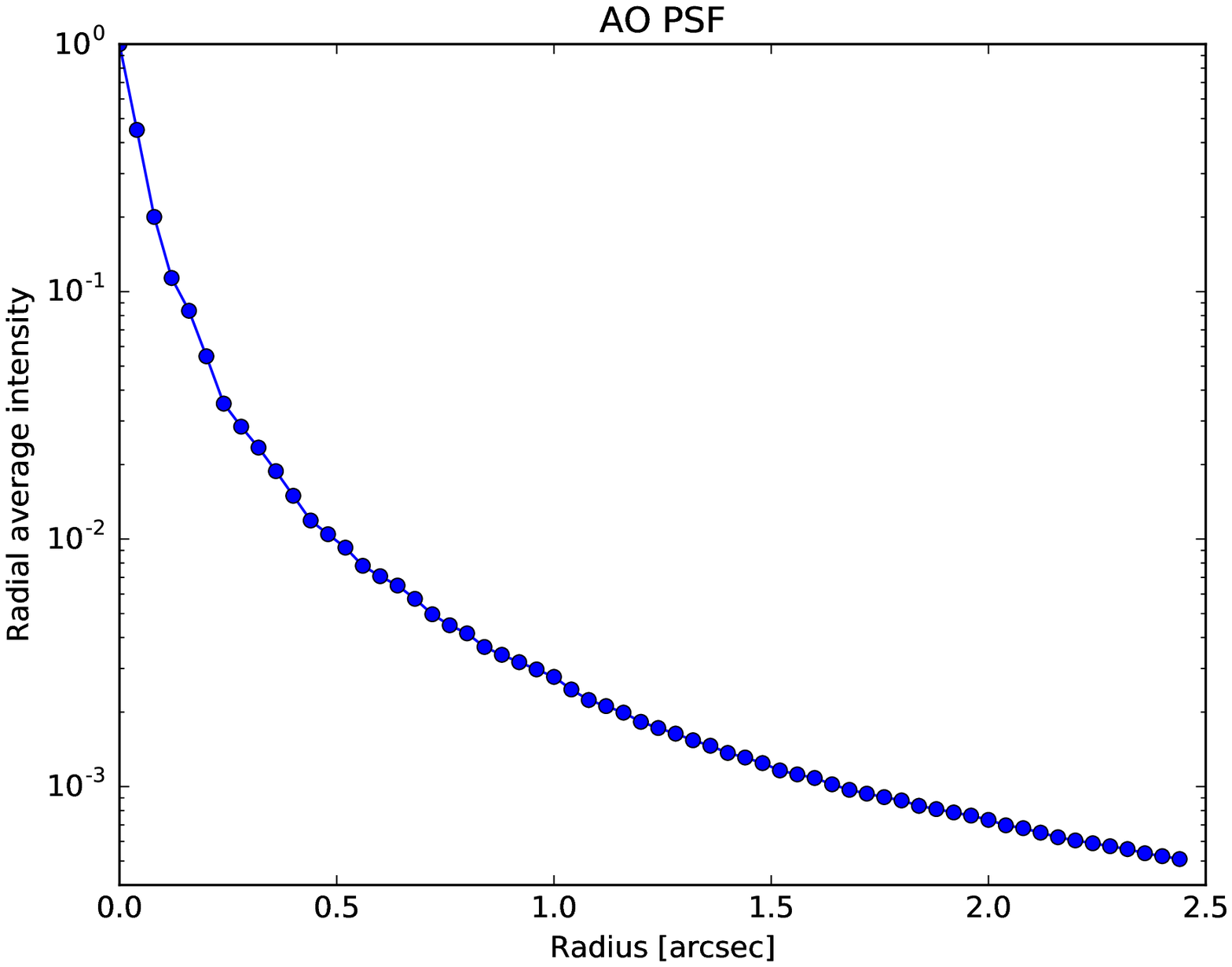}
  \caption{The left panel is the reconstructed AO PSF. The right panel is the radial average intensity of the PSF, 
    which shows the core plus its wings.}

  \label{fig:AO_PSF}
\end{figure*}


\subsection{Mock \#1: a sharp and rich structured PSF}
The mock \#1 image has $200 \times 200$ surface brightness pixels as 
constraints. The pixel size is $0.04''$. The simulated time delays in days 
relative to image B are: 
$\Delta t_{\rm AB}=1.5\pm1.5$, $\Delta t_{\rm CB}=-0.5\pm1.5$, $\Delta
t_{\rm DB}=90.5\pm1.5$.  

We follow the procedure described in Section \ref{sec:method} and Figure 
\ref{fig:diagram}. The reconstructions are shown in the second row of 
Figure \ref{fig:mock1_1}. To demonstrate the iterative processs visually, 
we show the process in Figure \ref{fig:mock1_2}. The first column shows 
each PSF correction grid in different iteration, the second column shows 
the cumulative PSF correction from iteration $1$ to iteration $18$, the third 
column is 
the PSF model at each iteration, and the right-most column shows the best 
fitting residuals with current PSF model. It is obvious that we get better 
and better normalized residuals as the iterative PSF corrections proceed. 
We follow Section \ref{sec:method:innerloopstop} and increase gradually the 
size of the PSF; the size of the final PSF is $85\times 85$ (which corresponds 
to $3.4''\times 3.4''$). However, since the PSF is very sharp in mock \#1, 
the PSF size with 
$19\times19$ 
 (which corresponds to $0.76''\times0.76''$) for the blurring 
matrix is enough. 
While $19\times19$ is sufficient for the extended source/lens light,
it is not for
the AGNs; $85\times 85$ is needed for describing the AGNs.

We try a series of source resolutions from coarse to fine, and the 
parameter constraints stabilize starting at $\sim$ $52\times52$ source pixels, 
corresponding to source pixel size of $\sim 0.045''$. 
In order to quantify the systematic uncertainty, we consider the following 
set of source resolutions: $52 \times 52$, $54 \times 54$, 
$56 \times 56$, $58 \times 58$, $60 \times 60$, and $62 \times 62$. 
We weight each choice of the source resolution equally\footnote{We weight the 
chains by the same weight because the source evidence are similar, and the 
lens parameterizations are the same.}, and combine the Markov chains together. 
The time delays are also reproduced by the model: for the 
various source resolutions, the total $\chi^2$ is $\sim 3$ for the three delays.
We demonstrate the important parameters for cosmography in 
the upper panel of
Figure 
\ref{fig:mock1_sample} 
(time-delay distance, external shear, radial slope of the main lens galaxy,
Einstein radii of the main galaxy and its satellite, total Einstein radius). 
The white dots represent the input values. The results show that we can 
recover the important parameters for cosmography. There is a strong degeneracy 
between the Einstein radii of the main galaxy and the satellite galaxy, as
expected since these two galaxies are both located within the arcs.
However, the effect on time-delay distance due to the presence of the
satellite is less than $1\%$ \citep{suyuEtal13}.
Despite the degeneracy, we can recover the total
Einstein radius within 1$\sigma$,
where the total Einstein radius, $\theta_{\rm {E,tot}}$, is defined by
\begin{equation}
\frac{\int_{0}^{\theta_{\rm E,tot}}\int_{0}^{2\pi}\kappa_{\rm tot}(\theta,\varphi)~d\varphi~d\theta}{\pi \theta_{\rm E,tot}^{2}}=1,
\end{equation} 
$\kappa_{\rm tot}$ is the total projected mass density including
the main galaxy and its satellite, and $\varphi$ is the polar angle on
the image plane. The total Einstein radius in here is only a circular approximation for the elliptical galaxy plus its satellite.

\begin{figure*}
\centering
\includegraphics*[scale=0.7]{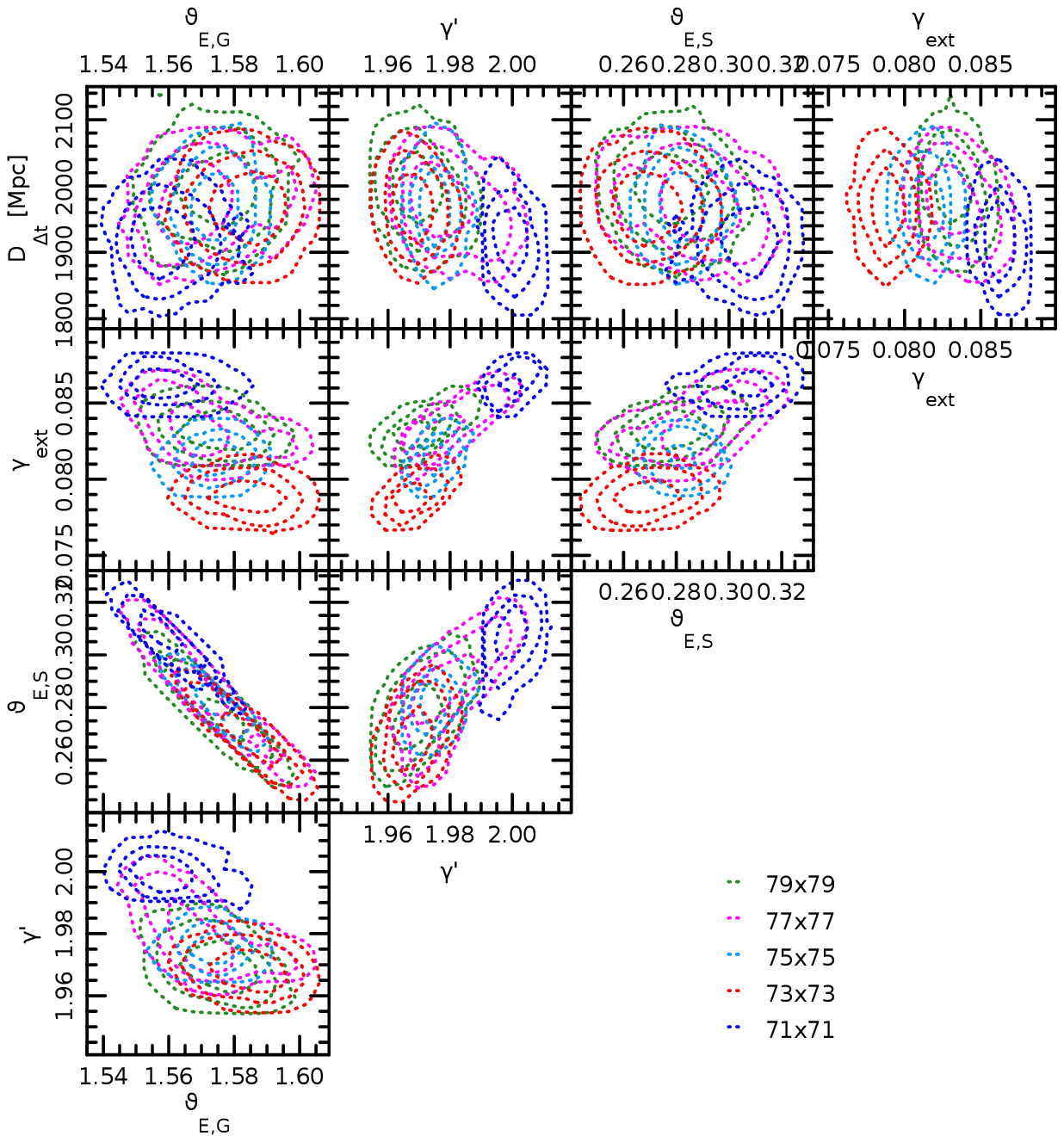}
\caption{Posterior of the key lens model parameters for \rxj\ and the
  time delays. We use the PSF size, $59\times59$, for convolution
  of the spatially extended lens and arcs.  We show the constraints
  from Markov chains of 
  different source resolutions: $71\times71$, $73\times73$,
  $75\times75$, $77\times77$, and $79\times79$. The contours
  mark the $68.3\%$, $95.4\%$, and $99.7\%$ credible regions
  for each source resolution.  The spread in the
  constraints from different chains allow us to quantify the 
  systematic uncertainty due to the pixelated source resolution.  
}
\label{fig:real_59}
\end{figure*}

\begin{figure*}
\centering
\includegraphics*[scale=0.7]{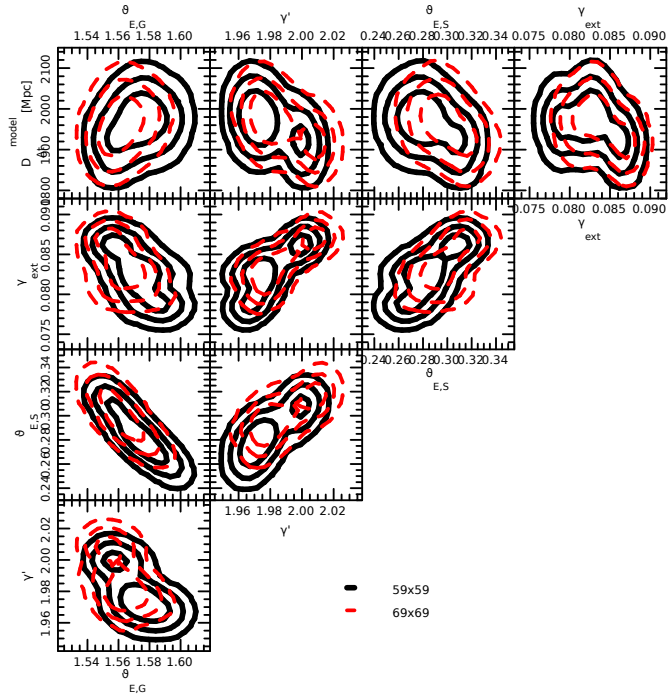}
\caption{
Posterior of the key lens model parameters for RXJ1131 and
  the time delays. We compare the PSF size, $59\times59$ and
  $69\times69$, for convolution of the spatially extended lens and
  arcs.  The constraints correspond to the combination of Markov
  chains of different source resolutions ($71\times71$, $73\times73$,
  $75\times75$, $77\times77$, and $79\times79$) in both PSF sizes. The
  contours mark the $68.3\%$, $95.4\%$, and $99.7\%$ credible
  regions. The constraints of the two PSF sizes are in
  good agreement, indicating that PSF sizes larger than
  $\sim59\times59$ are sufficient to 
  capture the PSF features for convolving the spatially extended
  images.
}
\label{fig:real_59and69}
\end{figure*}

\subsection{Mock \#2: a diffuse and smooth PSF}
The mock \#2 image has $300\times 300$ surface brightness pixels as constraints 
(the larger dimensions of the image are necessary
  for modeling the diffuse PSF). The pixel size and time delays are
the same as in  
mock \#1. The size of the final PSF is $127\times127$ (which corresponds to $5.08''\times 5.08''$). 
Since the PSF is very diffuse in mock \#2, the PSF size with $59 \times 59$ 
(which corresponds to $2.36''\times 2.36''$) 
for the blurring matrix is needed 
to convolve the spatially extended images.
We show the reconstruction in the fourth row of Figure \ref{fig:mock1_1}.

We also try a series of source resolutions from coarse to fine, and 
the parameter constraints stabilize starting at $\sim$ $59 \times
59$. To quantify systematic uncertainties due to source resolution, we consider the following set
of source resolutions: $59 \times 59$, $60 \times 60$, $61 \times 61$,
$62 \times 62$, and $63 \times 63$.  We also weight each source
resolution equally, and combine the Markov chains together. We
show the constraints on the same important parameters as mock \#1 
for cosmography in 
the lower panel of 
 Figure \ref{fig:mock1_sample}. The white dots
represent the input values. The results show that we can recover the 
important parameters for cosmography. 
Again, although we cannot recover the individual
Einstein radius due to the strong degeneracy between these two
Einstein radii, we can still recover the total Einstein radius.

We use the source-intensity-weighted regularization in the source 
reconstruction to prevent the source from fitting to the noise. 
The noise-overfitting problem is due to the fact that the outer region of 
the source plane is under-regularized. We do two tests which show its 
negligible impact on cosmographic inference: (1) We test it by changing the 
image covariance, ${\bf C}_{\text{D}}$, such that the uncertainties 
corresponding to low surface brightness areas are boosted (which is a 
similar effect as allowing the source to be more regularized at low surface 
brightness regions). The results show that the relative posteriors of 
lens/cosmological parameters are insensitive to 
such changes of
${\bf C}_{\text{D}}$ (hence the source regularization);
(2) We impose the source-intensity weighted regularization on the source 
plane, which can regularize more on the low surface brightness area on the 
source plane \citep[see e.g., ][for another type of regularization based on analytic profile]{Tagore14}.  Specifically, we obtain the first version of the source intensity distribution $\boldsymbol{s}_{\rm f}$ on a grid of pixels following the method of \citet{SuyuEtal06} with a constant regularization for all source pixels.  We then repeat the source reconstruction but with the regularization constant $\lambda$ scaled inversely proportional to $s_{\rm f}^{4}$, allowing high/low source intensity regions to be less/more regularized.
The relative posteriors between the different MCMC samples in 
the chains are the same between the uniform and source-intensity-weighted 
source regularizations. Furthermore, even with different source reconstrcution 
methods, the Einstein radius, which also plays an important role in 
cosmographic inference, is still robust.

\begin{figure*}
\centering
\includegraphics*[scale=0.49]{combo_AO_HST_2s+1_stats3.eps}
\includegraphics*[scale=0.32]{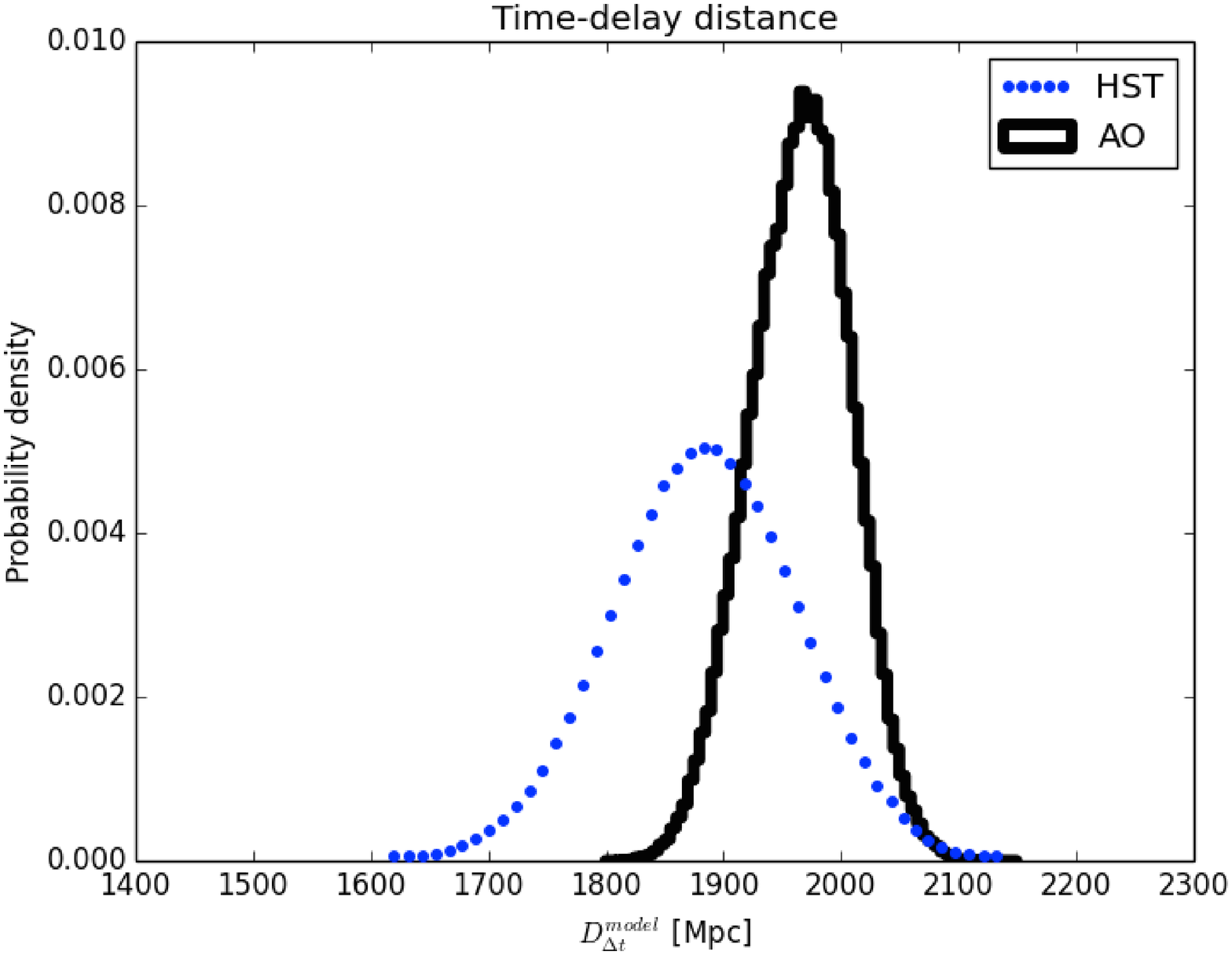}
\caption{
{\bf Left panel}: comparison of posterior of the key lens model parameters between AO
imaging (dashed) and \hst\ imaging (shades).  The AO constraints are
from the combination of both the $59\times 59$ and $69\times 69$
chains containing the series of source resolutions (e.g., Figure
\ref{fig:real_59and69} for $59\times 59$).  The contours/shades mark
the $68.3\%$, $95.4\%$, and $99.7\%$ credible regions.  The AO
constraints are consistent with the \hst\ constraints, and are in fact
$\sim50\%$ tighter on the modeled time-delay distance. {\bf Right panel}: PDFs for D$_{\Delta \text{t}}$, showing the constraints from \hst\ image and AO image.} 
\label{fig:AOHST}
\end{figure*}

\section{Real data modeling}
\label{sec:realdata}
We apply our newly-developed PSF reconstruction
method to the real AO imaging shown in Section \ref{sec:obs}, 
and use the time delays from \citet{TewesEtal13a}.
For the lens light, we use two S$\acute{\text{e}}$rsic profiles 
with common centroids and position angles for the main lens galaxy, 
and use one circular S$\acute{\text{e}}$rsic profile for the small 
satellite (whereas in the mock data in Section 
\ref{sec:blindtest} we describe the light of the satellite as a point source 
with PSF, $\PSF$). We find that, in this AO image, 4 concentric Gaussian 
profiles provide a good description of the initial global structure of the 
PSF\footnote{Due to
  unknown PSF, we do not have prior information on PSF. Thus, we test
  multiple concentric Gaussian profiles to fit the AGN. However, we
  find that the initial PSF model does not affect the final results which is 
  shown in Section \ref{sec:blindtest}, because the iterative method will 
  correct it in the end.}, 
which is the procedure we discussed in Section 
\ref{sec:method:initPSF:AGNlight}. For modeling the main lens mass, 
we use an elliptical symmetric distribution with 
power-law profile and an external shear which are described in Section 
\ref{sec:method:initPSF:arclight}; for modeling the mass
distribution of the satellite, we use
a pseudo-isothermal mass distribution. 

After we increase the PSF grid during the iterative
reconstruction scheme, the final PSF size is $127\times127$ 
(which corresponds to $5.08''\times5.08''$).
We try a series of source resolutions from coarse to fine and a series of PSF 
sizes for the blurring matrix from small to large. The parameter
constraints stabilize starting at $\sim 71\times71$ for the source resolution
and at $\sim 59\times59$ for the PSF size for the blurring matrix, corresponding
to a source pixel size of $\sim 0.05''$ and a PSF size of $2.36''\times 2.36''$.
Note, again, that while a PSF cutout of $59\times59$ is sufficient for
the extended source, the AGNs require a larger PSF grid of $127\times127$. 
We show the reconstructions of AO imaging in Figure \ref{fig:real_demo}
\footnote{We use the source-intensity-weighted regularization in the source 
reconstruction} and the reconstructed PSF in Figure \ref{fig:AO_PSF}. 
To quantify the systematic uncertainty, we show in Figure \ref{fig:real_59} 
the parameter constraints of different sizes of the source grid, $71\times 71$, 
$73\times 73$, $75\times 75$, $77\times 77$, and $79\times 79$, with the PSF 
size, $59\times59$, for the blurring matrix. 
After combining all the chains with different source resolutions, 
we overlap the contours from the $59\times59$ PSF with the contours from 
the $69\times69$ PSF (for the blurring matrix) in Figure
\ref{fig:real_59and69}; 
the results agree with each other within $1-\sigma$ uncertainty.

Since the PSF in \rxj\ AO imaging is similar to the PSF of mock \#2, 
the results from Figure \ref{fig:mock1_sample} provide a valuable reference.
Thus, note that the Einstein radii of
the main galaxy and the satellite galaxy inferred from the Keck AO
image are also degenerate with each other, as we saw in the case of mock \#2.

\begin{table}
\caption{Lens Model Parameter} 
\label{tab:parameters} 
\begin{center} 
\begin{tabular}{lll}  
\hline\hline 
Description & Parameter & Marginalized \\
&&or Optimized \\
&&Constraints\\
\hline 
Time-delay distance (Mpc)              & $D^{\rm model}_{\Delta t}$ & $1970^{+40}_{-43}$ \\ 
\hline
Lens mass distribution                & \\ 
\hline 
Centroid of G in $\theta_{1}$ (arcsec) & $\theta_{1,\rm G}$\tnote{$\clubsuit$} & $6.306^{+0.004}_{-0.008}$\\ 
Centroid of G in $\theta_{2}$ (arcsec) & $\theta_{2,\rm G}$ & $5.955^{+0.005}_{-0.005}$ \\ 
Axis ratio of G                       & $q_{\rm G}$ & $0.753^{+0.008}_{-0.007}$\\
Position angle of G ($\degree$)       &$\phi_{\rm G}$\tnote{$\spadesuit$} & $113.4^{+0.4}_{-0.5}$\\ 
Einstein radius of G (arcsec)         & $\theta_{\rm E,G}$ & $1.57^{+0.01}_{-0.01}$\\
Radial slope of G                     & $\gamma^{\prime}$ & $1.98^{+0.07}_{-0.02}$      \\
Centroid of S in $\theta_{1}$ (arcsec) & $\theta_{1,\rm S}$& $6.27^{+0.02}_{-0.03}$\\
Centroid of S in $\theta_{2}$ (arcsec) & $\theta_{2,\rm S}$& $6.56^{+0.01}_{-0.01}$\\
Einstein radius of S (arcsec)         & $\theta_{\rm E,S}$& $0.282^{+0.003}_{-0.003}$ \\
External shear strength               & $\gamma_{\rm ext}$& $0.083^{+0.003}_{-0.003}$\\
External shear angle ($\degree$)      & $\phi_{\rm ext}$& $93^{+1}_{-1}$\\
\hline
Lens light as S$\acute{\text{e}}$rsic profiles\\
\hline
Centroid of S in $\theta_{1}$ (arcsec) & $\theta_{1,\rm GL}$& $6.3052^{+0.0002}_{-0.0002}$\\
Centroid of S in $\theta_{2}$ (arcsec) & $\theta_{2,\rm GL}$& $6.0660^{+0.0002}_{-0.0002}$\\
Position angle of G ($\degree$)       &$\phi_{\rm GL}$ & $116.9^{+0.4}_{-0.4}$\\ 
Axis ratio of G1                      & $q_{\rm G}$& $0.912^{+0.004}_{-0.004}$\\
Amplitude of G1                       & $I_{\rm s,GL1}$\tnote{$\dagger$}&$1.47^{+0.02}_{-0.02}$ \\
Effective radius of G1 (arcsec)       & $R_{\rm eff,GL1}$&$2.37^{+0.01}_{-0.01}$\\
Index of G1                           & $n_{\text{s}\acute{\text{e}}\text{rsic},\rm GL1}$& $0.63^{+0.01}_{-0.01}$\\
Axis ratio of G2                      & $q_{\rm GL2}$& $0.867^{+0.002}_{-0.002}$\\
Amplitude of G2                       & $I_{\rm s,GL2}$& $18.1^{+0.3}_{-0.3}$\\
Effective radius of G2 (arcsec)       & $R_{\rm eff,GL2}$&$0.404^{+0.005}_{-0.005}$\\
Index of G2                           & $n_{\text{s}\acute{\text{e}}\text{rsic},\rm GL2}$&$1.97^{+0.02}_{-0.02}$\\
Centroid of S in $\theta_{1}$ (arcsec) & $\theta_{1,\rm SL}$& $6.210^{+0.001}_{-0.001}$\\
Centroid of S in $\theta_{2}$ (arcsec) & $\theta_{2,\rm SL}$& $6.605^{+0.001}_{-0.001}$\\
Axis ratio of S                       & $q_{\rm SL}$& $\equiv1$\\
Amplitude of S                        & $I_{\rm s,SL}$& $69^{+6}_{-6}$\\
Effective radius of S (arcsec)        & $R_{\rm eff,SL}$& $0.027^{+0.001}_{-0.001}$\\
Index of S                            & $n_{\text{s}\acute{\text{e}}\text{rsic},\rm SL}$&$0.42^{+0.04}_{-0.02}$\\
\hline
Lensed AGN light\\
\hline
Position of image A in $\theta_{1}$ (arcsec) & $\theta_{1,\rm A}$ & $4.256$\\
Position of image A in $\theta_{2}$ (arcsec) & $\theta_{2,\rm A}$ & $6.652$\\
Amplitude of image A                        & $a_{\rm A}$        & $21880$\\
Position of image B in $\theta_{1}$ (arcsec) & $\theta_{1,\rm B}$ & $4.288$\\
Position of image B in $\theta_{2}$ (arcsec) & $\theta_{2,\rm B}$ & $4.348$\\
Amplitude of image B                        & $a_{\rm B}$        & $38555$\\
Position of image C in $\theta_{1}$ (arcsec) & $\theta_{1,\rm C}$ & $4.871$\\
Position of image C in $\theta_{2}$ (arcsec) & $\theta_{2,\rm C}$ & $4.348$\\
Amplitude of image C                        & $a_{\rm C}$        & $11565$\\
Position of image D in $\theta_{1}$ (arcsec) & $\theta_{1,\rm D}$ & $7.378$\\
Position of image D in $\theta_{2}$ (arcsec) & $\theta_{2,\rm D}$ & $6.340$\\
Amplitude of image D                        & $a_{\rm D}$        & $3215$\\
\hline

\end{tabular} 
\begin{tablenotes}
  \item[$\clubsuit$]The reference of the position is in Figure \ref{fig:real_demo}.
  \item[$\spadesuit$]All the position angles are measured conterclockwise from positive $\theta_{2}$ (north).
  \item[$\dagger$]The amplitude is in equation (\ref{eq:lenslight0}).
\end{tablenotes}
\begin{tablenotes}[para,flushleft]
{\bf Note}: There are total 39 parameters that are optimized or sampled. The optimal parameters have little effect on the key parameters for cosmology (such as $D^{\rm model}_{\Delta t}$). For the lens light, two S$\acute{\text{e}}$rsic profiles with common centroid and position angle are used to describe the main lens galaxy G. They are denoted as G1 and G2 above. The source pixel parameters ($\s$) are marginalized and are thus not listed.
\end{tablenotes}
\end{center}
\end{table}

By using the same time delay meausurements from \citet{TewesEtal13a}
as in \citet{suyuEtal13}, 
we compare the results of modeling the AO image with the results of
modeling the \hst\ image from \citet{suyuEtal13}.\footnote{The mass model 
parameterization is the same as \citet{suyuEtal13} except for a slight difference in the definition of $\theta_{\rm E}$ due to ellipticity.  In this paper, we compare the $\theta_{\rm E}$ as defined in equation (\ref{eq:arclight-1}).  Thus the $\theta_{\rm E}$ shown in this paper is slightly different from that of \citet{suyuEtal13}.}  
We show the
comparison in Figure \ref{fig:AOHST} and list all the lens model parameters 
in Table \ref{tab:parameters}.  Except for
the highly degenerate Einstein radius of the main galaxy, other
important parameters are overlapping within 1-$\sigma$ uncertainty. 
Furthermore,
the constraint of time-delay distance by using AO imaging with
$0.045''$ resolution is tighter than the constraint of time-delay
distance by using \hst\ imaging with $0.09''$ by around 50$\%$.

For cosmographic measurement from time-delay lenses, we need to break
the mass-sheet degeneracy in gravitational lensing
\citep[e.g.,][]{FalcoEtal85, SchneiderSluse13, SchneiderSluse14,
  XuEtal15} that can change the modeled time-delay distance.  This
would involve considerations of mass profiles, lens stellar kinematics and external 
convergence \citep[e.g.,][]{TreuKoopmans02, BarnabeEtal11, suyuEtal13,
  SuyuEtal14} that are beyond the scope of this paper.  The focus of
this paper is to investigate the feasibility of AO imaging for follow
up.  As illustrated in Figure \ref{fig:AOHST}, AO
imaging together with our new PSF reconstruction technique (especially
of quad lens systems) is a competitive alternative to \hst\ imaging
for following up time-delay lenses for accurate lens modeling.

\section{Summary} 
\label{sec:summary}
In this paper we develop a new method, iterative PSF correction scheme, 
which can overcome the unknown PSF problem, 
to constrain cosmology by modeling the strong lensing AO 
imaging with time delays. 
We elaborate the procedures in Section \ref{sec:method} and draw an 
overall flow chart in Figure \ref{fig:diagram}. 

We test the method on two mock systems, mock \#1 (blindly) and mock \#2, 
which are created by using a sharp PSF and diffuse 
PSF, respectively, and apply this method to the 
high-resolution AO \rxj\ image taken with the Keck telescope 
as part of the SHARP AO observation. We draw the following conclusions.
\begin{itemize}
  \item We perform a blind test on mock \#1, which mimics the appearance
    of \rxj\ but with a sharp and richly 
structured PSF (based on a star observed with Keck's LGSAO). 
Afterward, we model the mock \#2, which is created by a diffuse PSF that 
is similar with the PSF in AO \rxj\ image, using the same strategy. 
The results show that the more 
diffuse PSF the AO imaging has, the larger the PSF is needed for 
representing the AGN; similarly, the larger the PSF for representing the
AGNs, the larger the PSF is needed for convolution of the 
spatially extended lens and arcs. By performing MCMC sampling, 
we can recover the important parameters for cosmography (time-delay 
distance, external shear, slope, and total Einstein radius of 
the main galaxy plus its satellite). Although we cannot recover the 
individual Einstein radius, 
the effect on time-delay distance due to the presence of the
satellite is less than $1\%$ \citep{suyuEtal13}.
  \item We model the AO \rxj\ image by the iterative PSF correction
    scheme. We compare the results of important parameters with the
    results from modeling the \hst\ imaging in
    \citet{suyuEtal13}. Except for the highly degenerate
    Einstein radius of the main galaxy, other important parameters for
    cosmography agree with each other within
    1-$\sigma$ (Figure \ref{fig:AOHST}). Furthermore, the constraint
    of time-delay distance by using AO imaging with $0.045''$
    resolution is tighter than the constraint of time-delay distance
    by using \hst\ imaging with $0.09''$ by around 50$\%$.
\end{itemize}

The iterative PSF reconstruction method that we have developed is
general and widely applicable to studies that require high-precision
PSF reconstruction from multiple nearby point sources in the field
(e.g., the search of faint companions of stars in star clusters).  For
the case of gravitational lens time delays, this method lifts the
restriction of using \hst\ strong lensing imaging,
and opens a new series of AO imaging data set to study cosmology. 
From the upcoming surveys, hundreds of new lenses are predicted 
to be discovered; this method not only can motivate more telescopes 
to be equipped with AO technology, but also facilitate the goal to reveal 
possible new physics by beating down the uncertainty on $H_{0}$ to
1$\%$ via strong lensing \citep{Suyuetal12b}.

\section*{Acknowledgments}

We thank Giuseppe Bono, James Chan, Thomas Lai, Anja von der Linden, Eric Linder, Phil Marshall, 
and David Spergel for the useful discussions. 
G.C.F.C.~and S.H.S.~are grateful to Bau-Ching
Hsieh for computing support on the SuMIRe computing cluster.
G.C.F.C.~and S.H.S.~acknowledge support from the Ministry of Science and 
Technology in Taiwan via grant MOST-103-2112-M-001-003-MY3.
LVEK is supported in part through an NWO-VICI career grant (project number 
639.043.308). CDF and DJL acknowledge support from NSF-AST-0909119.
The data presented herein were obtained at the W. M. Keck Observatory, which 
is operated as a scientific partnership among the California Institute of 
Technology, the University of California and the National Aeronautics and 
Space Administration. The Observatory was made possible by the generous 
financial support of the W. M. Keck Foundation. The authors wish to recognize 
and acknowledge the very significant cultural role and reverence that the 
summit of Mauna Kea has always had within the indigenous Hawaiian community. 
We are most fortunate to have the opportunity to conduct observations from 
this mountain.



\appendix

\section{Arc and AGN mask regions}
\label{app:a}
We show the three different mask regions, maskArcAGN (mArcAGN), maskArc (mArc),
and maskAGNcenter (mAc) in Figure \ref{fig:maskregion}. For modeling the
lens light in Section \ref{sec:method:initPSF:lenslight}, we mask out
the region which contains significant
arc light and AGN light in the left panel. For modeling the arc light in 
Section \ref{sec:method:initPSF:arclight}, we mask out the region with 
significant AGN light shown in the middle panel. 
For extracting the PSF correction, we show the residuals in the right panel 
(which is the image with the lens light, arc light, and AGN light subtracted). 
When the size of the correction grid is small such that the correction 
grids do not overlap other 
AGN center, we only need to mask out the area where it comes obviously 
from the host galaxy of AGN. 
For instance, if the background AGN has compact bright blobs in its host 
galaxy, 
due to the limit of the resolution on the source plane, the predicted arc 
cannot 
reconstruct the compact blobs,  so there are residuals around these compact
blobs on the image plane (shown in the right panel with red arrows). 
In order to prevent the correction grid from absorbing the light due to the 
resolution problem 
and adding non-PSF features into the PSF, we mask them out. When the 
correction grid is enlarged and covers other  
AGN centers, we need to mask out both regions (AGN centers and lensed 
compact blob).

\begin{figure*}
\centering
\includegraphics*[scale=0.55]{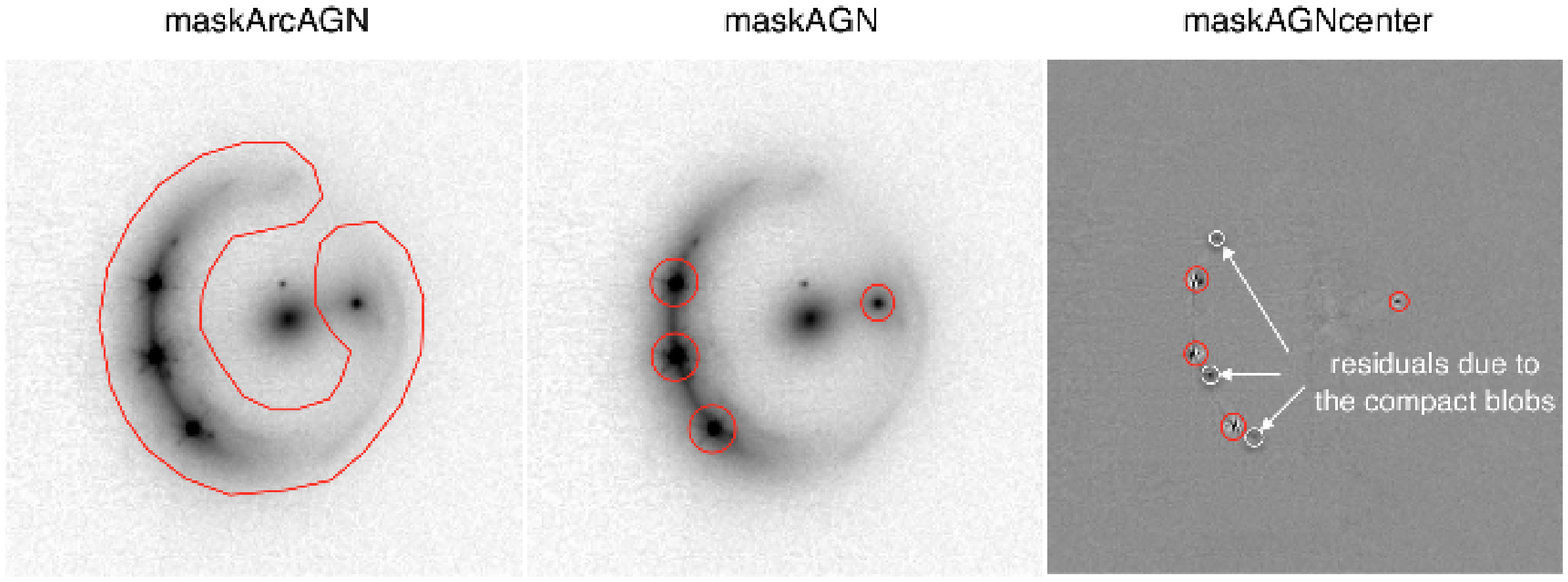}
\caption{The three different mask regions which are circled in red,
  and the white arrows indicate the special area which need to be masked
  out (that is, we boost the uncertainty in that region) while we
  extract the PSF correction. The
  left panel shows the maskArcAGN region for fitting the lens light,
  and the middle panel shows the maskAGN region for fitting the arc
  light.  When obtaining the PSF corrections, the white circles in the
  right panel need to be masked out when the PSF grid is small.  As
  we increase the PSF grid around each AGN image such that the grid
  contains other AGN images (shown in the right panel of Figure \ref{fig:pixgrid_overlap}), we mask out the red circles
  associated with these other AGN images and also the white circles.}
\label{fig:maskregion}
\end{figure*}

\begin{figure*}
\centering
\includegraphics*[scale=0.3]{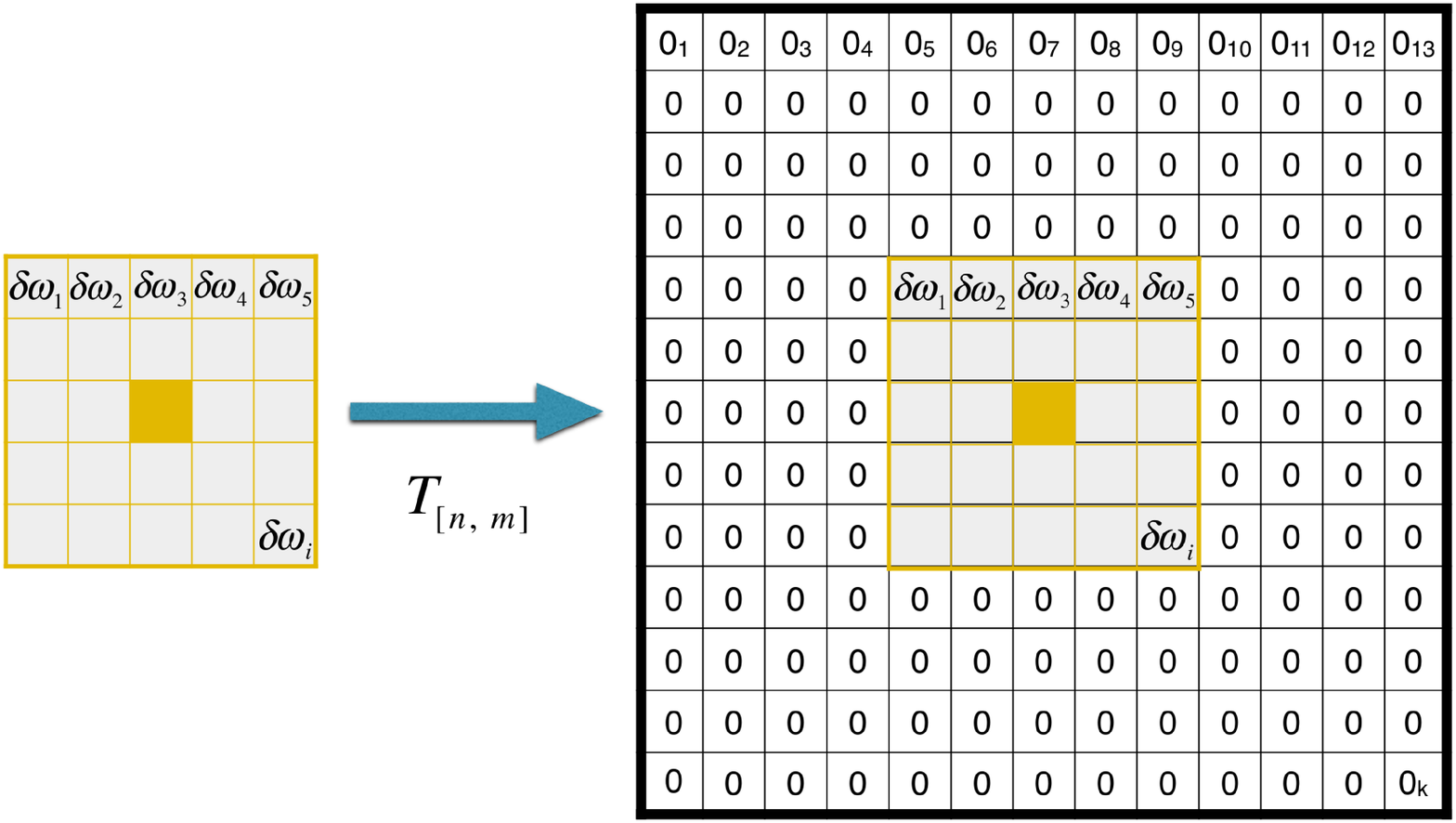}
\caption{
The matrix ${\Tvec}_{[\ol,\il]}$ for making
  $\delta\PSF_{[\ol,\il]}$the same length as $\PSF_{[\ol,\il]}$.  
The indices of $\delta\PSF_{i}$ and $0_{k}$ are for the pixels (rather
than the PSF correction iterations).  
  ${\Tvec}_{[\ol,\il]}$ is a matrix at the $\ol^{\rm th}$ outer loop
  and the $\il^{\rm th}$ inner loop.
}
\label{fig:matrixT}
\end{figure*}

\section{${\Tvec}_{[\ol,\il]}$ matrix}
\label{app:b}
Since $\delta\PSF_{[\ol,\il]}$ has different
length in each iteration of inner/outer loop ${[\ol,\il]}$, we use a
matrix ${\Tvec}_{[\ol,\il]}$ to make $\delta\PSF_{[\ol,\il]}$
the same length as $\PSF_{[\ol,\il]}$ by padding the two-dimensional
boundaries of the PSF correction grid with zeros, as illustrated
in Figure \ref{fig:matrixT}.





\bibliographystyle{mnras}
\bibliography{AOpaper_mnras} 

\bsp	
\label{lastpage}
\end{document}